\documentclass[aps,prl,twocolumn,superscriptaddress,longbibliography]{revtex4-2}
\usepackage{graphicx}
\usepackage{dcolumn}   
\usepackage{bm}        
\usepackage{amssymb}   
\usepackage{amsmath}
\usepackage[Symbol]{upgreek}
\usepackage{url}
\usepackage{layout}
\usepackage{color}
\usepackage{epsfig}
\usepackage{graphicx}
\usepackage{mathrsfs}\usepackage{bm}\usepackage{appendix}\usepackage{color}

\usepackage[thinlines]{easytable}
\usepackage{makecell}
\usepackage{tikz,pgf}
\usepackage{booktabs}
\usepackage{float}
\usepackage{subfigure}

\usepackage{cancel}

\usepackage{hyperref}
\hypersetup{colorlinks,allcolors=blue}

\begin{document}

\title{Possible Liquid-Nitrogen-Temperature Superconductivity Driven by Perpendicular Electric Field in the Single-Bilayer Film of La$_3$Ni$_2$O$_7$  at Ambient Pressure}
\author{Zhi-Yan Shao}
\thanks{These two authors contributed equally to this work.}
\affiliation{School of Physics, Beijing Institute of Technology, Beijing 100081, China}
\author{Jia-Heng Ji}
\thanks{These two authors contributed equally to this work.}
\affiliation{School of Physics, Beijing Institute of Technology, Beijing 100081, China}
\author{Congjun Wu}
\affiliation{New Cornerstone Science Laboratory, Department of Physics, School of Science, Westlake University, Hangzhou 310024, Zhejiang, China}
\affiliation{Institute for Theoretical Sciences, Westlake University, Hangzhou 310024, Zhejiang, China}
\affiliation{Key Laboratory for Quantum Materials of Zhejiang Province, School of Science, Westlake University, Hangzhou 310024, Zhejiang, China}
\affiliation{Institute of Natural Sciences, Westlake Institute for Advanced Study, Hangzhou 310024, Zhejiang, China}
\author{Dao-Xin Yao}
\affiliation{Guangdong Provincial Key Laboratory of Magnetoelectric Physics and Devices, State Key Laboratory of Optoelectronic Materials and Technologies, Center for Neutron Science and Technology, School of Physics, Sun Yat-Sen University, Guangzhou, 510275, China}
\author{Fan Yang}
\email{yangfan\_blg@bit.edu.cn}
\affiliation{School of Physics, Beijing Institute of Technology, Beijing 100081, China}

\begin{abstract}
\textbf{Abstract:} Recently, high-temperature superconductivity (HTSC) is found in the La$_3$Ni$_2$O$_7$/SrLaAlO$_4$ ultrathin film with critical temperature $T_c$ above the McMillan limit at ambient pressure (AP). It is eager to enhance $T_c$ of La$_3$Ni$_2$O$_7$ at AP. We propose that a perpendicular electric field strongly enhances $T_c$ in the single-bilayer film of La$_3$Ni$_2$O$_7$ at AP. Under electric field, the layer with lower potential energy will accept electrons flowing from the other layer to fill in the Ni-$3d_{x^2-y^2}$ orbitals, as the nearly half-filled Ni-$3d_{z^2}$ orbital cannot accommodate more electrons. With the enhancement of the filling fraction in the $3d_{x^2-y^2}$ orbitals in this layer, the interlayer $s$-wave pairing is suppressed, but the intralayer $d$-wave pairing in this layer is strongly enhanced. We numerically verify this idea and yield that an imposed voltage of about $0.1\sim0.2$ volt between layers is enough to realize liquid-nitrogen-temperature HTSC in this single bilayer at AP. Our results appeal for experimental verification.
\end{abstract}

\maketitle

\section{Introduction}
The discovery of superconductivity (SC) with critical temperature $T_c$ above the boiling point of liquid nitrogen ($\approx 77 \text{ K}$) in the pressurized La$_3$Ni$_2$O$_7$~\cite{Wang2023LNO, YuanHQ2023LNO,Wang2023LNOb,wang2023LNOpoly,wang2023la2prnio7,zhang2023pressure,zhou2023evidence,wang2024bulk,li2024pressure} has attracted great interests \cite{Fukamachi2001,khasanov2024pressure,chen2024evidence,dan2024spin,chen2024electronic,Wang2022LNO,Kakoi2024,xie2024neutron,gupta2024anisotropic,ren2024resolving,feng2024unaltered,meng2024density,fan2024tunn,xu2024pressure,LI2024distinct,liu2024electronic,yashima2025microscopic,khasanov2025oxygen,yang2024orbital,wang2023structure,cui2023strain,Li2024ele,li2024distinguishing,zhou2024revealing,wang2025chemical,Chen2024poly,Dong2024vis,Li2024design,puphal2024unconven,zhu2024superconductivity,zhang2023superconductivity,huang2024signature,li2023trilayer,zhang2020intertwined,xu2024origin,du2024correlated,li2019superconductivity,lee2023linear,nomura2022superconductivity,gu2022superconductivity,sui2023rno,YaoDX2023,Dagotto2023,cao2023flat,zhang2023structural,huang2023impurity,geisler2023structural,rhodes2023structural,zhang2023la3ni2o6,yuan2023trilayer,li2024la3,geisler2024optical,li2017fermiology,wang2024non,chen2024tri,ZhangGM2023DMRG, Werner2023,shilenko2023correlated,WuWei2023charge,chen2023critical,ouyang2023hund,heier2023competing,wang2024electronic,botzel2024theory,WangQH2023,YangF2023,lechermann2023,Kuroki2023,HuJP2023,lu2023bilayertJ,oh2023type2,liao2023electron,qu2023bilayer,Yi_Feng2023,jiang2023high,zhang2023trends,qin2023high,tian2023correlation,jiang2023pressure,lu2023sc,kitamine2023,luo2023high,zhang2023strong,pan2023rno,sakakibara2023La4Ni3O10,lange2023mixedtj,yang2023strong,lange2023feshbach,kaneko2023pair,fan2023sc,wu2024deconfined,zhang2024prediction,zhang2024s,Yang2024effective,zhang2024electronic,yang2024decom,ryee2024quenched,Lu2024interplay,Ouyang2024absence,labollita2024,zhang2024emergent,Leonov2024Electronic,labollita2024assessing,ni2025spin,Yi2024nature,LaBollita2024electronic,jiang2024theory,chen2024non,zhang2024magnetic,lin2024magnetic,qin2024intertwined,Leonov2024Electronicc,liu2024growth,liu2024decoupling,tian2024spin,yin2025spmep,PhysRevB.111.104505,kaneko2025tj,Ji2025StrongCouplingLimit, Wang_2025,haque2025dft,shi2025theoretical, Ko2024signature,zhou2025ambient,liu2025superconductivity,yue2025correlated,li2025photoemission,bhatt2025resolving,shao2025band,shi2025effect,le2025landscape,Daoxin_Yao2025,wang2025electronic,huang2025spm,geisler2025electronic}.  This discovery has sparked the exploration of high-temperature SC (HTSC) in Ruddlesden-Popper phase multilayer nickelates, resulting in the discovery of SC in the pressurized La$_4$Ni$_3$O$_{10}$~\cite{zhu2024superconductivity,zhang2023superconductivity,huang2024signature,li2023trilayer,zhang2020intertwined,xu2024origin,du2024correlated}, which together with the previously synthesized infinite-layer nickelates Nd$_{1-x}$Sr$_x$NiO$_2$~\cite{li2019superconductivity,lee2023linear,nomura2022superconductivity,gu2022superconductivity} have established a new family of SC other than cuprates and iron-based superconductors. 
However, the high pressure (HP) circumstance not only strongly hinders the experimental detection of the samples but also brings difficulties in the application of SC in industry. Very recently, the La\textsubscript{3}Ni\textsubscript{2}O\textsubscript{7} ultrathin film with a few layers of unit cell grown on the SrLaAlO\textsubscript{4} (SLAO) substrate has been grown by two different groups independently and SC with $T_c$ above the McMillan limit ($\approx$40 K) has been detected at ambient pressure (AP)\cite{Ko2024signature,zhou2025ambient,liu2025superconductivity}, allowing various experimental investigation of the pairing mechanism in this material, attracting a lot of interests \cite{yue2025correlated,li2025photoemission,bhatt2025resolving,shao2025band,shi2025effect,le2025landscape,Daoxin_Yao2025,wang2025electronic,huang2025spm,geisler2025electronic}. It is now eager to enhance the $T_c$ of this material at AP. Here we propose a viable approach to realize $T_c$ above the boiling point of liquid nitrogen in the La\textsubscript{3}Ni\textsubscript{2}O\textsubscript{7} single-bilayer film at AP. 

Presently, the pairing mechanism in the La$_3$Ni$_2$O$_7$, either in the bulk material under HP~\cite{WangQH2023,YangF2023,lechermann2023,Kuroki2023,HuJP2023,lu2023bilayertJ,oh2023type2,liao2023electron,qu2023bilayer,Yi_Feng2023,jiang2023high,zhang2023trends,qin2023high,tian2023correlation,jiang2023pressure,lu2023sc,kitamine2023,luo2023high,zhang2023strong,pan2023rno,sakakibara2023La4Ni3O10,lange2023mixedtj,yang2023strong,lange2023feshbach,kaneko2023pair,fan2023sc,wu2024deconfined,zhang2024electronic,yang2024decom,ryee2024quenched,Lu2024interplay,Ouyang2024absence,yin2025spmep,PhysRevB.111.104505,kaneko2025tj,Ji2025StrongCouplingLimit} or in the ultrathin film at AP~\cite{yue2025correlated,shao2025band,le2025landscape,huang2025spm}, is still under debate. Density-functional-theory (DFT) based first-principle calculations have suggested that the low-energy orbitals are mainly Ni-$3d_{z^2}$ and $3d_{x^2-y^2}$, which are nearly half- and quarter- filled~\cite{sui2023rno,YaoDX2023,Dagotto2023,cao2023flat,zhang2023structural,huang2023impurity,geisler2023structural,rhodes2023structural,zhang2023la3ni2o6,geisler2024optical}. Various experiments have revealed the strongly-correlated characteristic of the material~\cite{chen2024electronic,Kakoi2024,fan2024tunn,LI2024distinct,liu2024electronic,yang2024orbital,Li2024ele,li2024distinguishing}. Particularly, the optical study reveals significant reduction of the electron kinetic energy which places the system in the proximity of the Mott phase \cite{liu2024electronic}; the angle-resolved photoemission spectroscopy uncovers strong band renormalization caused by electron correlation~\cite{yang2024orbital}; the linearly temperature-dependent resistivity suggests ``strange-metal'' behavior~\cite{YuanHQ2023LNO}. Therefore, we take a strong-coupling view of the system. Under the strong Hubbard repulsion, the nearly half-filled $3d_{z^2}$ electrons can almost be viewed as localized spins. Therefore, the main carrier of SC should be the $3d_{x^2-y^2}$ electrons, which subject to the in-plane superexchange interaction just mimic the 50\% hole-doped cuprates. However, it is a problem how HTSC can emerge under such a high doping level. The key physics lies in the important role played by the $3d_{z^2}$ orbitals. Through strong interlayer perpendicular superexchange, the $3d_{z^2}$ electrons form interlayer pairing. The interlayer perpendicular superexchange or the interlayer pairing of the $3d_{z^2}$ electrons can be transmitted to the $3d_{x^2-y^2}$ electrons through the Hund's rule~\cite{lu2023bilayertJ,oh2023type2,qu2023bilayer,tian2023correlation,zhang2023strong,pan2023rno,lange2023mixedtj,yang2023strong,lange2023feshbach,kaneko2023pair,wu2024deconfined,Lu2024interplay,Ouyang2024absence} or the nearest-neighbor (NN) hybridization~\cite{ZhangGM2023DMRG,Yi_Feng2023,qin2023high,tian2023correlation,luo2023high,kaneko2023pair,yang2024decom,Lu2024interplay,Ouyang2024absence} or both. Under such view, the role of pressure in enhancing the $T_c$ lies in the enhancement of the interlayer perpendicular superexchange, the inter-orbital hybridization, or both. 

\begin{figure}[htbp]
    \centering
    \includegraphics[width=1\linewidth]{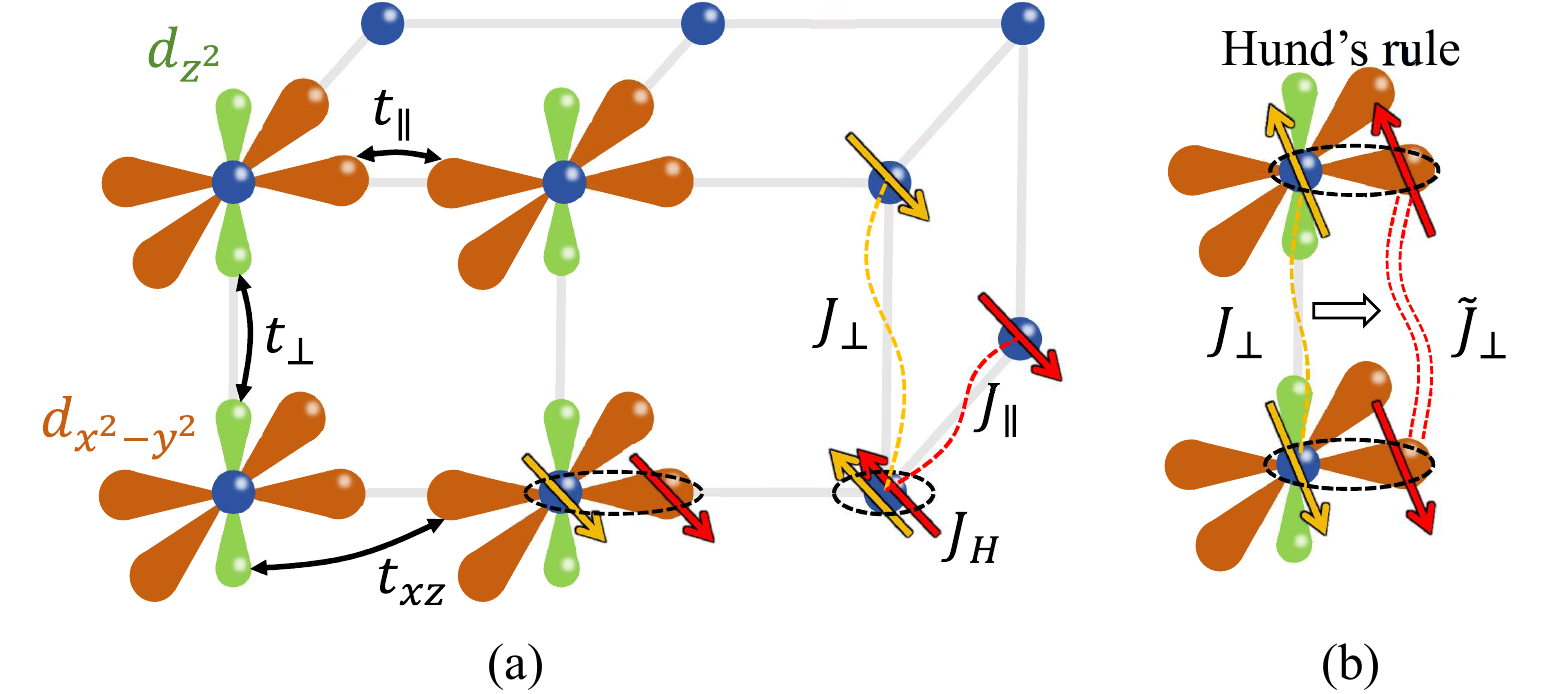}
    \caption{Schematic diagrams of the model. (a) Schematic diagram for the dominant hopping integrals and superexchange interactions between the $E_g$ orbitals in La$_3$Ni$_2$O$_7$.  (b) Schematic diagram illustrating that the Hund's rule coupling transmits the interlayer perpendicular superexchange interaction $J_\perp$ between the $3d_{z^2}$ orbitals to the effective one $\tilde{J}_\perp$ between the $3d_{x^2-y^2}$ orbitals.} \label{fig_SD}
\end{figure}

In this work, we propose an alternative approach to realize HTSC with $T_c$ above the boiling point of liquid nitrogen in the ultrathin film of La$_3$Ni$_2$O$_7$ at AP. 
Here we consider the thinnest limit, i.e. a single bilayer film of La$_3$Ni$_2$O$_7$, and realize the goal by introducing charge transfer with a perpendicular electric field, which let the electrons flow from the high-energy layer to the low-energy layer, similar to the mechanism for the spontaneous charge transfer in oxide heterostructures \cite{PhysRevB.95.205131,PhysRevLett.123.117201,PhysRevB.98.155103,PhysRevB.108.054441,mondal2023modulation,doi:10.1021/acs.nanolett.4c02087}.
The external electric field based approach avoids introducing disorder as in chemical doping~\cite{jiang2021electronic} or exhibiting orbital selectivity based on symmetry~\cite{sohn2024observation}, demonstrating exceptional performance in the field of twisted multilayer graphene materials~\cite{cao2020tunable,zhang2024observation,zhou2025gate}.
We can impose a perpendicular electric field, say pointing upward, in this single bilayer, so that electrons from the top layer will flow to the bottom layer. These electrons will fill the $3d_{x^2-y^2}$ orbitals in the bottom layer as the nearly half-filled $3d_{z^2}$ orbitals there cannot accommodate more electrons. The enhancement of the bottom-layer $3d_{x^2-y^2}$ electron number will first suppress the interlayer $s$-wave SC due to mismatch of the electron numbers between the two layers, 
similar with the case in which an imposed Zeeman field acting on the spin leads to mismatch of the electron numbers between the two spin species and thus suppresses singlet pairing, 
and then promptly lead to the intra-bottom-layer $d$-wave SC with strongly enhanced $T_c$. To test this idea, we have performed a combined simplified single orbital study and a comprehensive two orbital study, which consistently yield that a voltage of experimentally achievable levels (around $0.1\sim 0.2$ volt predicted by the mean-field calculations) between the two layers is enough to induce $d$-wave SC with $T_c$ above the boiling point of liquid nitrogen in the bottom layer. Intriguingly, the $d$-wave SC carried by the bottom layer $3d_{x^2-y^2}$ electrons coexists with the interlayer $s$-wave pseudo gap carried by the $3d_{z^2}$ electrons in the mixing ratio of $1:\mathrm{i}$, breaking time-reversal symmetry. Our proposal potentially provides a viable approach to realize HTSC with $T_c$ above the boiling point of liquid nitrogen in the single bilayer film of La$_3$Ni$_2$O$_7$.

\section{Results}
\subsection{Consideration and a Simplified Study}
\begin{figure}
    \centering
    \includegraphics[width=1\linewidth]{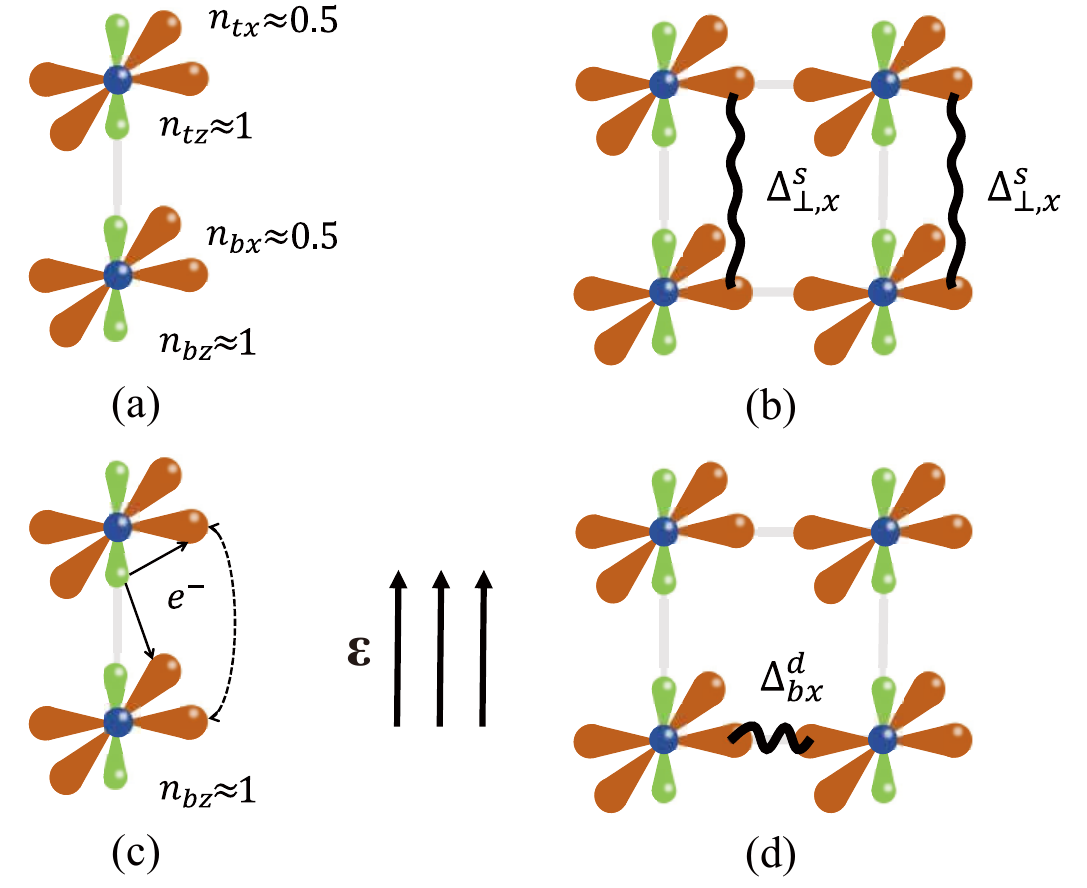}
    \caption{Schematic diagrams of particle number and pairing configuration before and after introducing perpendicular electric field. (a) Particle numbers of the four $E_g$ orbitals within an unit cell without electric field. (b) The dominant pairing configuration for (a). (c) Schematic diagram showing how the electrons flow under the perpendicular electric field $\bm{\upvarepsilon}$ pointing upward. (d) The dominant pairing configuration for (c).}
    \label{fig_idea}
\end{figure}

The La\textsubscript{3}Ni\textsubscript{2}O\textsubscript{7} ultrathin film grown on the SLAO substrate form a bilayer square lattice \cite{bhatt2025resolving,yue2025correlated}. 
As illustrated in Fig.~\ref{fig_SD} (a), the leading hopping integrals are the interlayer hopping of the $3d_{z^2}$ electrons $t_{\perp}$ and the intralayer NN hopping of the $3d_{x^2-y^2}$ electrons $t_{\parallel}$. Under strong Hubbard $U$, these hopping terms can induce the effective superexchange $J_{\perp}$ and $J_{\parallel}$ through $J\approx \frac{4t^2}{U}$. Under the Hund's rule coupling $J_H$, the spins of the two orbitals are inclined to be parallel aligned, as illustrated in Fig.~\ref{fig_SD} (b), which partly transmits the interlayer perpendicular superexchange $J_{\perp}$ between the $3d_{z^2}$ orbitals to the $3d_{x^2-y^2}$ orbitals as $\tilde{J}_{\perp}=\alpha J_{\perp}$ with $\alpha\in (0,1)$ describing the efficiency of this process and related to the strength of Hund's coupling $J_H$. In addition, there exists intralayer NN- hybridization $t_{xz}$ between the two orbitals. As shown in Fig.~\ref{fig_idea} (a, b), the nearly quarter-filled $3d_{x^2-y^2}$ electrons subject to $J_{\parallel}$ and $\tilde{J}_{\perp}$ form interlayer-dominant pairing~\cite{lu2023bilayertJ}. 

Now let us turn on the upward electric field $\bm{\upvarepsilon}$, forcing electrons downward, as shown in Fig.~\ref{fig_idea} (c, d). In the top layer, since the $3d_{z^2}$ orbitals host larger density of state (DOS) than the $3d_{x^2-y^2}$ orbitals, they will donate more electrons. Most of these donated electrons will fill the $3d_{x^2-y^2}$ orbitals in the bottom layer, as the nearly half-filled $3d_{z^2}$ orbitals there cannot accommodate more electrons. A minority of the donated electrons can also be accepted by the top-layer $3d_{x^2-y^2}$ orbitals.  

\begin{figure}[htbp]
    \centering
    \includegraphics[width=1\linewidth]{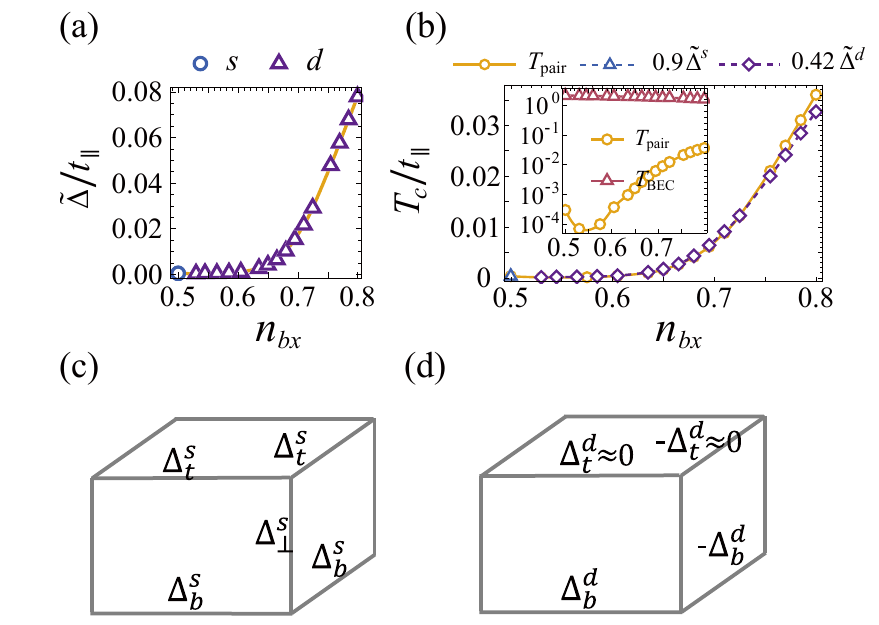}
    \caption{The SBMF results for the single-orbital model. (a) The pairing amplitude $\tilde{\Delta}$ (in unit of $t_{\parallel}$) as function of the bottom-layer particle number per site $n_{bx}$. Different pairing symmetries are distinguished by color.  (b) The $T_c$ as function of $n_{bx}$, in comparison with $0.42\tilde{\Delta}$ for the $d$-wave and $0.9\tilde{\Delta}$ for the $s$-wave regime. Inset: the spinon pairing temperature $T_{\mathrm{pair}}$ and the holon condensation temperature $T_{\mathrm{BEC}}$ as function of $n_{bx}$. In (a,b), we set $J_{\parallel}=0.4t_{\parallel}$ and $\tilde{J}_{\perp}=(1-\delta_{tz})\times1.3J_{\parallel}$.  (c)-(d) The pairing configurations of the $s$-wave and $d$-wave, respectively.} \label{fig_gap}
\end{figure}

Even with doped holes under $\bm{\upvarepsilon}$, the top-layer $3d_{z^2}$ electrons cannot carry SC: Firstly, lacking pairing interaction, they cannot form intralayer pairing. Secondly, although they can pair with the localized bottom-layer $3d_{z^2}$ electrons, such pairs cannot coherently move, only resulting in the pseudo-gap. Therefore, the SC under $\bm{\upvarepsilon}$ can only be carried by the $3d_{x^2-y^2}$ orbitals. As the filling fractions of the $3d_{x^2-y^2}$ orbitals in the two layers are different, their Fermi levels are relatively shift, leading to mismatch of their Fermi surfaces (FSs), which will suppress their interlayer pairing. 
Here the perpendicular electric field acts as a ``pseudo-Zeeman field'' acting on the layer index, just like the Zeeman field acting on the spin degree of freedom. 
The bottom-layer $3d_{x^2-y^2}$ orbitals will form $d$-wave SC, mimicsing the cuprates, as shown in Fig.~\ref{fig_idea} (d). When $\bm{\upvarepsilon}$ is strong enough so that the filling fraction of the bottom-layer $3d_{x^2-y^2}$ orbitals is near that of the optimal doped cuprates, $d$-wave HTSC with strongly enhanced $T_c$ will be achieved in this layer, and the top layer will also acquire SC below $T_c$ through proximity.

Based on the above general consideration, we first conduct the following simplified model study including only the $3d_{x^2-y^2}$-orbital, with the $3d_{z^2}$ orbital only viewed as a source which tunes the total electron number. The following widely adopted single $3d_{x^2-y^2}$-orbital bilayer $t-J-J_{\perp}$ model~\cite{lu2023bilayertJ,oh2023type2,qu2023bilayer,zhang2023strong,pan2023rno,lange2023mixedtj,yang2023strong,wu2024deconfined} is adopted,
\begin{eqnarray} \label{eq_tJJperpe}
H=&-&t_{\parallel} \sum_{\left\langle{}i,j\right\rangle,\mu,\sigma} \hat{P} \left( c^{\dag}_{i\mu\sigma}c_{j\mu\sigma} + \mathrm{h.c.} \right) \hat{P} + \sum_{i,\mu} \epsilon_{\mu} n_{i\mu} \nonumber\\&+& J_{\parallel} \sum_{\left\langle{}i,j\right\rangle,\mu} \mathbf{S}_{i\mu} \cdot \mathbf{S}_{j\mu} +  \tilde{J}_{\perp}\sum_{i} \mathbf{S}_{it} \cdot \mathbf{S}_{ib}.
\end{eqnarray}
Here $c^{\dag}_{i\mu\sigma}$ creates an electron at site $i$ in the layer $\mu$ (=top ($t$)/bottom ($b$)) with spin $\sigma$, $\hat{P}$ is a projection operator projecting out the double occupancy of all site, and $n_{i\mu}$ or $\mathbf{S}_{i\mu}$ denote the corresponding electron number or spin operator. Only NN- bond $\langle i,j\rangle$ is considered in the summation.  The $\epsilon_{\mu}$ is introduced to control the filling fractions of the two layers under $\bm{\upvarepsilon}$, with $\epsilon_{t}-\epsilon_{b}=\varepsilon$. However, as the total particle number of the $d_{x^2-y^2}$ electrons under given $\bm{\upvarepsilon}$ is unknown, we have to assume the ratio $r:1$ between the electron number flowing from the $3d_{z^2}$ orbitals and that flowing from $3d_{x^2-y^2}$ orbitals in the top layer when solving the model with the standard slave-boson mean-field (SBMF) theory \cite{kotliar1988}
, which demonstrates exceptional performance for La$_3$Ni$_2$O$_7$ in previous studies~\cite{lu2023bilayertJ,oh2023type2} that is qualitatively consistent with experimental data~\cite{Wang2023LNO} and theoretical studies using other numerical methods (like DMRG)~\cite{qu2023bilayer,zhang2023strong, yang2023strong}.
Due to reason of DOS, we assume this ratio to be $2:1$, with details provided in the Supplementary Information (SI)~\cite{SM}. Nevertheless, the concrete value of this ratio turns out not to obviously affect the results (see the SI~\cite{SM}). The filling fractions are fixed under this assumption in the SBMF study. To capture the quantum fluctuation beyond mean-field, the density matrix renormalization group (DMRG)~\cite{white1993dmrg,weng1999dmrg} method is also employed, whose results are qualitatively consistent with those of the SBMF study, indicating that the SBMF theory can capture the main features of this system. See details in the \textbf{METHODS} and SI~\cite{SM}.



{We set $t_{\parallel}=1$ as the energy unit and $J_{\parallel}=0.4t_{\parallel}$ in our study. $\tilde{J}_{\perp} = (1-\delta_{tz})\times1.3J_{\parallel}$ is applied in our SBMF study. The results are shown in Fig.~\ref{fig_gap}. Fig.~\ref{fig_gap}~(a) shows the amplitude and symmetry of the ground-state pairing gap as function of the bottom-layer $3d_{x^2-y^2}$ electron number $n_{bx}$, whose value enhances with $\bm{\upvarepsilon}$. It is shown that when $\bm{\upvarepsilon}$ or $n_{bx}$ enhances, the pairing amplitude $\tilde{\Delta}$ decays first and then increases. 
When $n_{bx}=0.5$, the ground state is confirmed to be interlayer $s$-wave SC by comparing the energies of states with different symmetries (See SI \cite{SM} for more details).
Then the $s$-wave pairing is suppressed by the enhancement of $\varepsilon$ or $n_{bx}$ because of the mismatch of the FSs of the two layers caused by $\varepsilon$, similar to the case of a singlet pairing state placed within a pair-breaking Zeeman field. Therefore, it is also possible that this ``pseudo Zeeman field'' can drive pair density wave (PDW), just like that the real Zeeman field can drive the Fulde-Ferrell-Larkin-Ovchinnikov state. 
When $n_{bx} \geq 0.53$, the ground state is an intralayer $d$-wave SC, with the dominant pairing limited in the bottom layer. It is inspiring that with the enhancement of $n_{bx}$ in this regime, the $\tilde{\Delta}$ enhances promptly, similar to the case in the overdoped cuprates, wherein the enhancement of the filling fraction promptly enhances the pairing strength. The pairing configurations of the two different pairing symmetries are illustrated in Fig.~\ref{fig_gap}~(c-d).

The $T_c\sim n_{bx}$ is shown in Fig.~\ref{fig_gap}~(b). In the SBMF theory, the $T_c$ is given as the lower one between the spinon-pairing temperature $T_{\text{pair}}$ and the holon-BEC temperature $T_{\text{BEC}}$. The inset of Fig.~\ref{fig_gap}~(b) displays $T_{\text{BEC}}\gg T_{\text{pair}}$, rendering $T_c=T_{\text{pair}}$ in the considered $n_{bx}$ regime. Note that the $T_c$ here is in the sense of Kosterlitz-Thouless transition. A comparison between Fig.~\ref{fig_gap}~(b) and (a) suggests that $T_c$ scales with $\tilde{\Delta}$, which is more clear when the $T_c\sim n_{bx}$ is well fitted by $0.42\tilde{\Delta}\sim n_{bx}$ for the $d$-wave and $0.9\tilde{\Delta}\sim n_{bx}$ for the $s$-wave in Fig.~\ref{fig_gap}~(b), consistent with the Bardeen-Cooper-Schrieffer (BCS) theory. Inspiringly, for $n_{bx}\ge 0.75$, the $T_c\gtrsim 0.02t_{\parallel}\approx 80$ K, suggesting the HTSC in the liquid nitrogen temperature range.

On the above, we have adopted $\tilde{J}_{\perp}=\alpha J_{\perp}(1-\delta_{tz})$ with $\alpha=1$, where $\delta_{tz}$ denotes the hole density of the top-$3d_{z^2}$ orbital. For the reduced $\alpha$, only the low-$n_{bx}$ regime accommodating the interlayer $s$-wave pairing in Fig.~\ref{fig_gap}~(a, b) shrinks but the high-$n_{bx}$ regime accommodating the intralayer $d$-wave SC is not affected because the intralayer pairing is blind to $\tilde{J}_{\perp}$.  Furthermore, assuming different ratios between the changes of the filling fractions of the two top-layer $E_g$ orbitals turns out to yield similar results when expressed as functions of $n_{bx}$, as the dominant pairing under strong $\bm{\upvarepsilon}$ is the intra-bottom-layer pairing, which is blind to the filling fraction of the top layer. See the SI for details~\cite{SM}.

We have further employed the DMRG approach, which can capture the quantum fluctuation beyond mean-field, to compute the ground state of Hamiltonian (\ref{eq_tJJperpe}) under different electric fields $\varepsilon$ and the transferred electron-doping levels of the $d_{x^2-y^2}$ orbitals $\delta = n_{tx}+n_{bx}-1$. For $\varepsilon=0$, we have $\delta=0$. When $\varepsilon$ increases, it drives electrons from $d_{z^2}$-orbitals in the top layer to $d_{x^2-y^2}$-orbitals in both layers, increasing $\delta$. However, as the exact relationship between $\varepsilon$ and $\delta$ is unclear, we set them as two independent variables in our DMRG study. The parameters $t_{\parallel}$ and $J_{\parallel}$ take the same values as the ones in the SBMF study while $\tilde{J}_{\perp}=0.8J_{\parallel}$ is adopted in the DMRG study. To characterize the pairing symmetry and strength, we analyze the interlayer pairing correlation function $\Phi^\perp(r)$ and the intra-bottom-layer pairing correlation function $\Phi^{\parallel}_{b}(r)$. More details are provided in \textbf{METHODS}.

\begin{figure}[htbp]
    \centering
    \includegraphics[width=0.99\linewidth]{1orb_DMRG_main.pdf} 
    \caption{The DMRG results. (a) The $\delta-\varepsilon$ phase diagram of the ground state. The red region corresponds to the $s$-wave pairing and the blue region to the $d$-wave pairing.  (b)-(c) The absolute value of the intra-bottom-layer pairing correlation functions $|\Phi^{\parallel}_b(r)|$ under different electric fields $\varepsilon=0, 0.4t_\parallel, 0.8t_\parallel, 1.2t_\parallel, 1.6t_\parallel$ for $\delta=0$ in (b) and $\delta=1/16$ in (c). (d)-(e) $|\Phi^{\parallel}_b(r)|$ for different transferred $d_{x^2-y^2}$-electron-doping levels $\delta=0,1/16,1/8$ under $\varepsilon=0.4t_\parallel$ in (d) and $\varepsilon=0.8t_\parallel$ in (e). The algebraic decay exponents $K_\text{SC}$ are written in the four figures as well, reflecting the decay rate of the pairing correlation function with spatial distance, negatively correlated with the corresponding pairing strength. In (a-e), $\delta$ and $\varepsilon$ are set as independent variables, since their exact relationship is unclear.}\label{fig_dmrg}
\end{figure}

Fig. \ref{fig_dmrg} (a) shows the pairing phase diagram with respect to $\delta$ ($=0, 1/16, 1/8$) and $\varepsilon$ ($\in [0, 1.6t_\parallel]$). Fig. \ref{fig_dmrg} (b) shows the absolute value of the intra-bottom-layer pairing correlation functions $|\Phi^{\parallel}_b(r)|$ under different electric fields $\varepsilon=0, 0.4t_\parallel, 0.8t_\parallel, 1.2t_\parallel, 1.6t_\parallel$ for $\delta=0$, and the results for $\delta=1/16$ are presented in Fig. \ref{fig_dmrg} (c). It turns out that $|\Phi^{\parallel}_b(r)|$ exhibits algebraic decay under an non-zero external electric field with the decaying power exponent to be $K_{\text{SC}}$, i.e. $|\Phi^{\parallel}_b(r)|\propto r^{-K_{\text{SC}}}$ for large enough $r$, implying the presence of pairing within the bottom layers.
Fig. \ref{fig_dmrg} (d) and (e) depict $|\Phi^{\parallel}_b(r)|$ for different transferred $d_{x^2-y^2}$-electron-doping levels $\delta=0,1/16,1/8$ under $\varepsilon=0.4t_\parallel$ in (d) and $\varepsilon=0.8t_\parallel$ in (e).
All the algebraic decay exponents $K_\text{SC}$ are provided accordingly.

The results indicate that (i) With the enhancement of the perpendicular electric field $\varepsilon$, and hence the transferred $d_{x^2-y^2}$-electron-doping level $\delta$, the pairing symmetry changes from interlayer $s$-wave to intra-bottom-layer $d$-wave (The criterion of the pairing symmetry is provided in \textbf{METHODS}); (ii) For all the transferred $d_{x^2-y^2}$-electron-doping levels $\delta$ tested, the enhancement of the perpendicular electric field $\varepsilon$ leads to a reduction of $K_{\text{SC}}$, suggesting the enhancement of the intra-bottom-layer pairing; (iii) Under all the perpendicular electric field strengths $\varepsilon$ tested, the enhancement of the transferred $d_{x^2-y^2}$-electron-doping level $\delta$ leads to a reduction of $K_{\text{SC}}$, suggesting the enhancement of the intra-bottom-layer pairing. From (ii) and (iii), it is clear that the enhancement of $\varepsilon$ and hence $\delta$ will significantly enhance the intra-bottom-layer pairing. These results are qualitatively consistent with those of the SBMF study. More results are given in the SI \cite{SM}. 

Besides, we study the effect of interlayer Coulomb interaction. Our results show that the interlayer Coulomb interaction slightly promotes charge transfer between layers and the intra-bottom-layer pairing, while suppressing the interlayer pairing. See SI \cite{SM} for details. 

\subsection{The comprehensive two-orbital study}
The above simplified single-orbital study has drawbacks: We cannot determine the relationship between the electron-doping of the $d_{x^2-y^2}$ orbitals and the electric field. In the SBMF study, we have to assume the ratio between the changes of the filling fractions of the two top-layer $E_g$ orbitals. In addition, we do not know how the neglected $3d_{z^2}$ orbital degree of freedom affects the pairing nature. To settle these puzzles, we conduct a comprehensive two-orbital model~\cite{Lu2024interplay} to study with,
\begin{widetext}
\begin{eqnarray}\label{eq_H2}
     H = &-& t_\parallel\sum_{\langle i,j\rangle,\mu} \hat{P} \left(c^\dag_{i\mu x\sigma}c_{j\mu x\sigma} + \mathrm{h.c.}\right) \hat{P} - t_\perp\sum_i \hat{P} \left(c^\dag_{itz\sigma}c_{ibz\sigma} + \mathrm{h.c.}\right) \hat{P} - t_{xz} \sum_{\langle i,j\rangle\mu} \hat{P} \left(c^\dag_{i\mu x\sigma}c_{j\mu z\sigma}+(z\leftrightarrow x)+\mathrm{h.c.}\right) \hat{P} \nonumber\\
    &+& J_\parallel\sum_{\langle i,j\rangle\mu}\mathbf{S}_{i\mu x}\cdot\mathbf{S}_{j\mu x} + J_\perp\sum_{i}\mathbf{S}_{itz}\cdot\mathbf{S}_{ibz} + \tilde{J}_\perp\sum_{i}\mathbf{S}_{it x}\cdot\mathbf{S}_{ib x}+\epsilon_z\sum_{i\mu\sigma}n_{i\mu z\sigma} + \epsilon_x\sum_{i\mu\sigma}n_{i\mu x\sigma} \nonumber\\
    &+&\frac{\varepsilon}{2} \sum_{i\alpha\sigma} n_{it\alpha\sigma} - \frac{\varepsilon}{2} \sum_{i\alpha\sigma}n_{ib\alpha\sigma}.
\end{eqnarray}
\end{widetext}
The operators $c_{i\mu \alpha\sigma}$, $n_{i\mu \alpha}$, $\mathbf{S}_{i\mu \alpha}$ take the same meanings as those in model (\ref{eq_tJJperpe}) except for an extra index $\alpha=x/z$ labeling the orbital, and $\hat{P}$ is a projection operator projecting out the double occupancy in the same orbital of all sites. Note that $\mathbf{S}_{i\mu \alpha}$ for each orbital is spin-$\frac{1}{2}$ operator. $\epsilon_\alpha$  denotes the on-site energy of the orbital $\alpha$.
We adpot the tight-binding (TB) parameters reported in Ref. \cite{Daoxin_Yao2025}, i.e. $t_{\parallel}=0.445\text{ eV}$, $t_{xz}=0.221\text{ eV}$, $t_{\perp}=0.503\text{ eV}$ and $\epsilon_x-\epsilon_z=0.367\text{ eV}$. 
The superexchange interactions are obtained through $J_{\parallel}\approx 4t_{\parallel}^2/U$ and $J_{\perp}\approx4t_{\perp}^2/U$, with $U=10t_{\parallel}$. Finally $\varepsilon$ denotes the voltage between the two layers. 
Here, due to the weak super-exchange interaction between the $d_{z^2}$ orbitals in the layer, we do not consider this term in our model. In addition, the Hund's coupling $J_H$ of La$_3$Ni$_2$O$_7$ is generally considered to be in the range of 0.7 eV to 1 eV in past studies \cite{lechermann2023,ouyang2023hund,WuWei2023charge}, which only slightly larger than the largest hopping parameter $t_{\perp}=0.503~\mathrm{eV}$ and thus does not satisfy the premise of the Schriffer-Wolf transformation or perturbation theory, we do not apply it here.
More details are provided in \textbf{METHODS}.

Our SBMF results of Eq. (\ref{eq_H2}) (see \textbf{METHODS} and the SI~\cite{SM}) are shown in Fig.~\ref{fig_2orb_gap}. Fig.~\ref{fig_2orb_gap}(a) shows the $\varepsilon$-dependence of the hole densities $\delta_{\mu\alpha}$. Obviously, the $\delta_{tz}$ enhances obviously with $\varepsilon$, suggesting that the top-$3d_{z^2}$ orbital is donating electrons. 
These donated electrons flow to the $3d_{x^2-y^2}$ orbitals in both layers, with more of them flowing to the bottom layer when $\varepsilon>0.1\text{ eV}$ while about half of them flow to the bottom layer when $\varepsilon\leq0.1\text{ eV}$.  
Fig.~\ref{fig_2orb_gap}(b) shows the $\varepsilon$-dependence of the pairing symmetry and the pairing gap amplitude of the bottom-layer $3d_{x^2-y^2}$ orbital. At low $\varepsilon\le 0.03$ eV, the pairing symmetry is $s$-wave, whose pairing configuration is shown in Fig.~\ref{fig_2orb_gap}(c), wherein the $3d_{z^2}$-orbital form interlayer $s$-wave pseudo-gap, while the $3d_{x^2-y^2}$ orbital form $s$-wave SC with coexisting intralayer and interlayer pairing. 
In this regime the interlayer pairing is suppressed by the enhancement of $\varepsilon$ while the intralayer pairing is enhanced. When $\varepsilon=0$, the interlayer pairing gap is the largest. When $\varepsilon$ is about $0.01\sim0.03\text{ eV}$, the intralayer pairing gap is slightly larger than the interlayer pairing gap.
When $\varepsilon>0.03$ eV, the pairing symmetry is $d$($3d_{x^2-y^2}$)+i$s$($3d_{z^2}$), whose pairing configuration is shown in Fig.~\ref{fig_2orb_gap}(d). In this state, the $3d_{z^2}$ orbital form interlayer $s$-wave pseudo-gap, while the bottom-layer $3d_{x^2-y^2}$ orbital form intralayer $d$-wave SC. When $\varepsilon$ enhances in this regime, the pairing amplitude for the $d$-wave part enhances promptly. For $\varepsilon > 0.13$ eV, the pairing amplitude can go beyond $0.02$ eV. Then from the relation $T_c\approx 0.42\tilde{\Delta}$ for the $d$-wave SC illustrated in Fig.~\ref{fig_gap}(b), we have got HTSC with $T_c\gtrsim 80$ K!

The result shown in Fig.~\ref{fig_2orb_gap}(b) for the comprehensive two-orbital study and that shown in Fig.~\ref{fig_gap}(b) for the simplified one-orbital study look similar, except that in Fig.~\ref{fig_2orb_gap}(b) the result is expressed as function of the directly controllable quantity $\varepsilon$. Actually, if we replace the $x$-axis of Fig.~\ref{fig_2orb_gap}(b) by the calculated $n_{bx}=1-\delta_{bx}$, the resulting curve nearly coincides with Fig.~\ref{fig_gap}(b), particularly in the large-$n_{bx}$ regime, see the SI~\cite{SM}. The main reason for such similarity lies in that under strong $\bm{\upvarepsilon}$, the dominant superconducting pairing is the intra-bottom-layer $3d_{x^2-y^2}$-orbital pairing, which is insensitive to the $3d_{z^2}$ orbital. The main new information obtained in the two-orbital study lies in that the $3d_{z^2}$ orbital form interlayer $s$-wave pseudo-gap which is mixed with the intra-bottom-layer $d$-wave HTSC of the $3d_{x^2-y^2}$ orbital in the ratio of $1:\mathrm{i}$, as shown in Fig.~\ref{fig_2orb_gap}(d). This state breaks time-reversal symmetry, although the experimentally detected superconducting gap is the standard $d$-wave gap of the $3d_{x^2-y^2}$ orbital. This intriguing result is left for experimental verification. 

\begin{figure}[htbp]
    \centering
    \includegraphics[width=0.95\linewidth]{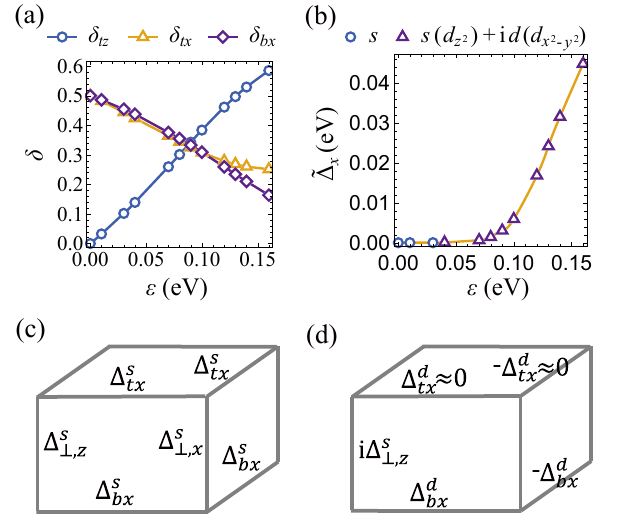}
    \caption{The SBMF results for the two-orbital model. (a) The hole densities $\delta_{\mu\alpha}$ for the three orbitals as functions of the strength of the electric field $\varepsilon$. (b) The pairing gap amplitude of the bottom-layer $3d_{x^2-y^2}$-orbital as function of $\varepsilon$. (c)-(d) The pairing configurations of the $s$-wave and the $d$($3d_{x^2-y^2}$)+i$s$($3d_{z^2}$)-wave, respectively.} \label{fig_2orb_gap}
\end{figure}

\section{Discussion}
In conclusion, we propose that an imposed strong perpendicular electric field can drive HTSC with $T_c$ above the boiling point of liquid nitrogen in the single-bilayer film of La$_3$Ni$_2$O$_7$ at AP. The reason lies in that under the strong electric field, the electrons in the layer with higher potential energy will flow to the layer with lower potential energy, to fill the $3d_{x^2-y^2}$ orbitals in the latter layer. 
When the imposed electric field is weak, it acts as the ``pseudo-Zeeman field'' operating on the layer index which supresses the interlayer SC, possibly inducing the PDW state. 
With considerably enhanced filling fraction, the $3d_{x^2-y^2}$ electrons in that layer just mimic the cuprates, which form intralayer $d$-wave HTSC with strongly enhanced $T_c$.  Our combined one-orbital and two-orbital studies consistently verify this idea. 

Presently, while different groups have provided slightly different TB parameters for the La$_3$Ni$_2$O$_7$ ultrathin film grown on SLAO substrate at AP, we have just adopted one set of these TB parameters to perform our calculations. However, the strong-coupling calculations performed here do not seriously rely on the accurate values of these parameters, because the main physics here is clear and simple. Actually, the well consistency between the result of the comprehensive two-orbital study and those of the simplified one-orbital studies with assuming different input conditions just verifies the robustness of our conclusion.  

Moreover, we want to emphasize that the essence of introducing the perpendicular electric field is breaking the symmetry of the two layers by making their filling fractions different to each other. Actually, the filling fractions of different NiO layers in the La\textsubscript{3}Ni\textsubscript{2}O\textsubscript{7} ultrathin film grown on the SLAO substrate may different from each other because of the existence of the substrate on one side of the film. This can be considered as effective electric field. Thus, our work provides a possible way to understand the high $T_c$ of the La\textsubscript{3}Ni\textsubscript{2}O\textsubscript{7} ultrathin film grown on the SLAO substrate.

\section{Methods}
\subsection{The one-orbital model}
The SBMF theory is used to solve the one-orbital model~(\ref{eq_tJJperpe}). In the SBMF approach, the superexchange terms are decomposed in $\chi-\Delta$ channel, e.g. $\mathbf{S}_{it}\cdot\mathbf{S}_{ib}=-\frac{3}{8}\left(\left\langle\chi^{\perp\dag}\right\rangle \chi^\perp_{i} + \mathrm{h.c.} + \left\langle\Delta^{\perp\dag}\right\rangle \Delta^\perp_{i} + \mathrm{h.c.}\right)$, $\chi$ and $\Delta$ represents hopping and pairing operators respectively. These MF parameters are further solved in a self-consistent manner. The specific steps can be referenced from prior work \cite{lu2023bilayertJ,kotliar1988,Lu2024interplay} and SI \cite{SM}.

We also employ the DMRG method \cite{white1993dmrg,weng1999dmrg} to solve the ground state of the Hamiltonian~(\ref{eq_tJJperpe}) as a comparison for the SBMF approach. The tensor libraries TensorKit \cite{jutho_2024_13950435} and FiniteMPS \cite{Li_FiniteMPS_jl_2024} provide an implementation of the required symmetry \cite{WEICHSELBAUM20122972,PhysRevResearch.2.023385}. We study the model on a $2\times 2\times L_x$ lattice with the open boundary conditions in the $x$ direction and choose $L_x=64$ for calculations. The matrix product state is constructed as shown in Fig. \ref{fig_mps}. We keep up to $D=12000$ $\mathrm{U}(1)_\text{charge}\times \mathrm{SU}(2)_\text{spin}$ multiplets in DMRG simulations and ensure the convergence accuracy of $10^{-6}$.

\begin{figure}[htbp]
    \centering
    \includegraphics[width=0.95\linewidth]{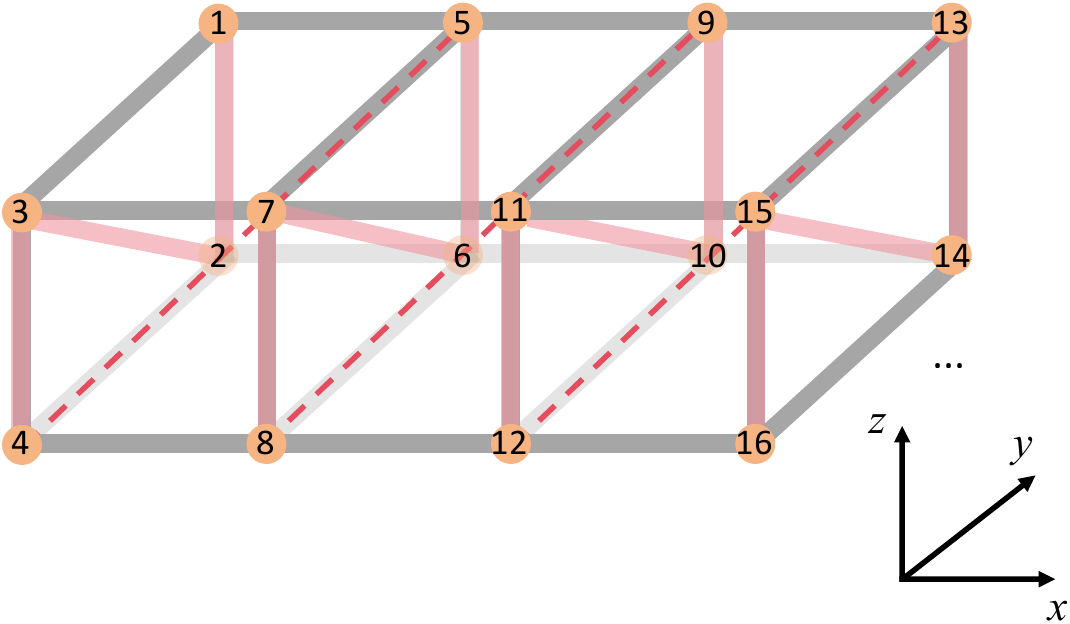}
    \caption{Illustration of the zigzag path in DMRG calculation.} \label{fig_mps}
\end{figure}

The interlayer and intralayer singlet pairing operators take the form of
\begin{equation}
\begin{aligned}
    &\Delta^{\perp\dag}_{i} = \frac{1}{\sqrt{2}}\left(c^{\dag}_{it\uparrow}c^{\dag}_{ib\downarrow}-c^{\dag}_{it\downarrow}c^{\dag}_{ib\uparrow}\right),\\
    &\Delta^{\parallel\dag}_{i\mu} \equiv \Delta^{\mathbf{x}\dag}_{i\mu}= \frac{1}{\sqrt{2}}\left(c^{\dag}_{i\mu\uparrow}c^{\dag}_{i+\mathbf{x},\mu\downarrow}-c^{\dag}_{i\mu\downarrow}c^{\dag}_{i+\mathbf{x},\mu\uparrow}\right),\\
    &\Delta^{\mathbf{y}\dag}_{i\mu} = \frac{1}{\sqrt{2}}\left(c^{\dag}_{i\mu\uparrow}c^{\dag}_{i+\mathbf{y},\mu\downarrow}-c^{\dag}_{i\mu\downarrow}c^{\dag}_{i+\mathbf{y},\mu\uparrow}\right).
\end{aligned}
\end{equation}
Here, the subscripts $i+\mathbf{x}$($i+\mathbf{y}$) represent the NN site of the site $\mathbf{i}$ in the $x$($y$) direction. Fig. \ref{fig_pairOp} shows how the singlet pairing operators are defined.

\begin{figure}[htbp]
    \centering
    \includegraphics[width=0.95\linewidth]{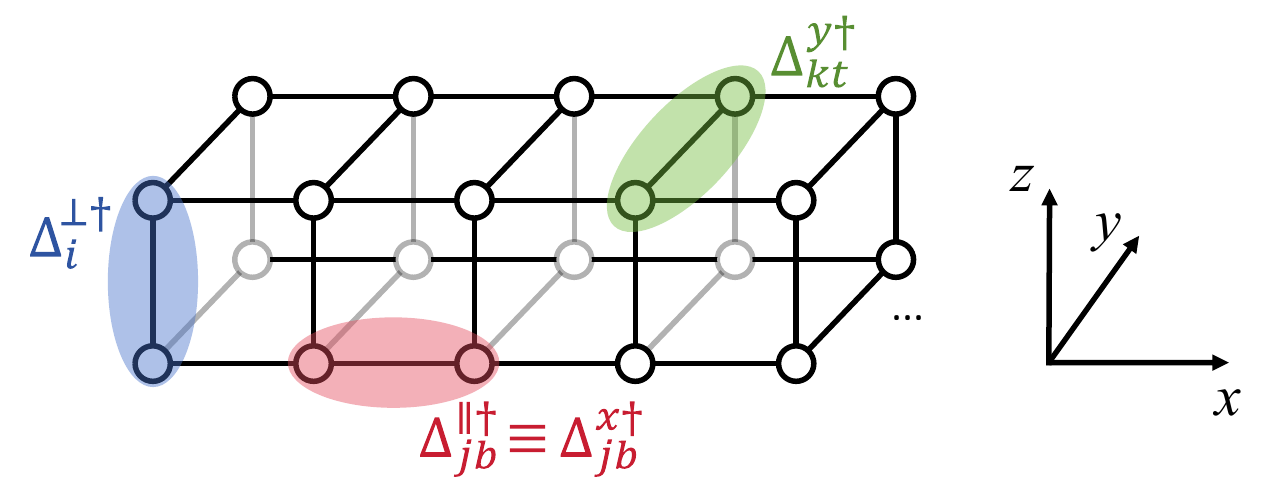}
    \caption{Illustration of the singlet pairing operators $\Delta^{\perp\dagger}_{i}$, $\Delta^{\parallel\dagger}_{i\mu}$ and $\Delta^{\mathbf{y}\dagger}_{i\mu}$.} \label{fig_pairOp}
\end{figure}

The considered correlation functions are defined as follow
\begin{equation}
\begin{aligned}
    &\Phi^{\perp}(r) = \left\langle \Delta^{\perp\dag}_{i} \Delta^{\perp}_{j} \right\rangle,\\
    &\Phi^{\parallel}_{\mu}(r) \equiv \Phi^{\mathbf{xx}}_\mu(r) =  \left\langle\Delta_{i\mu}^{\parallel\dag} \Delta^{\parallel}_{j\mu} \right\rangle,\\
    &\Phi^{\mathbf{xy}}_{\mu}(r) = \left\langle\Delta_{i\mu}^{\parallel\dag} \Delta^{\mathbf{y}}_{j\mu} \right\rangle,
\end{aligned}
\end{equation}
where $r=|\mathbf{i}-\mathbf{j}|$ is the distance between the sites $i$ and $j$. 

For a pairing channel whose absolute value of correlation function decays algebraically with distance, following the form $r^{-K_\text{SC}}$, the decay exponent $K_\text{SC}$ is associated with the Luttinger parameter specific to the channel. $K_{SC}<2$ signals a divergent superconducting susceptibility in that channel. The channel with the lowest $K_\text{SC}$ value is considered to dominate the pairing behavior.

The dominant pairing channel is related to pairing symmetry. For the case where interlayer pairing dominates, the pairing symmetry is restricted to $s$-wave pairing; while for the case where intralayer pairing in the bottom-layer dominates, we determine the pairing symmetry by the sign function $\mathrm{sgn}\left[\Phi^{\parallel}_{b}(r) \Phi^{\mathbf{xy}}_b(r)\right]$. If $\mathrm{sgn}\left[\Phi^{\parallel}_{b}(r) \Phi^{\mathbf{xy}}_b(r)\right]=-1$ holds for all $r$, the ground state can be identified as the $d$-wave pairing SC state. See SI~\cite{SM} for more details on DMRG.

\subsection{The two-orbital model}
Here we provide more technique details for the SBMF study on the two-orbital model~(\ref{eq_H2}). The electron operator is decomposed as $c^{\dag}_{i\mu\alpha\sigma}=f^{\dag}_{i\mu\alpha\sigma}b_{i\mu\alpha}$, where $f$ is spinon operator and $b$ is holon operator. Since we have found that $T_{\mathrm{BEC}} \gg T_{\mathrm{pair}}$ in the considered $n_{bx}$ regime and $T_{\mathrm{pair}}$ is proportional to the zero-temperature spinon pairing gap, we can get the critical temperature of superconductivity only by calculating the ground-state spinon pairing gap. Thus we only consider the spinon Hamiltonian at zero temperature. The superexchange term is also decomposed in $\chi-\Delta$ channel. The spinon Hamiltonian is described as
\begin{equation}\label{eq_MF2}
    \begin{aligned}
        H_{\mathrm{spinon}} = &- t_\parallel\sum_{\langle i,j\rangle,\mu}\delta_{\mu x}\left(f^\dag_{i\mu x\sigma}f_{j\mu x\sigma} + \mathrm{h.c.}\right)\\
        &- t_{xz}\sqrt{\delta_{tx}\delta_{tz}} \sum_{\langle i,j\rangle} \left(f^\dag_{itx\sigma}f_{jtz\sigma} + f^\dag_{itz\sigma}f_{jtx\sigma} + \mathrm{h.c.}\right)\\
        &- \frac{3}{8} J_\parallel\sum_{\langle i,j\rangle\mu}\left(\chi_{ij,\mu x}^\dag\left\langle\chi_{\mu x}\right\rangle + \mathrm{h.c.} - \left\langle\chi_{\mu x}^\dag\right\rangle\left\langle\chi_{\mu x}\right\rangle \right)\\
        &- \frac{3}{8} J_\parallel\sum_{\langle i,j\rangle\mu}\left(\Delta_{ij,\mu x}^\dag\left\langle\Delta_{\mu x}\right\rangle + \mathrm{h.c.} - \left\langle\Delta_{\mu x}^\dag\right\rangle\left\langle\Delta_{\mu x}\right\rangle\right)\\
        &-\frac{3}{8} J_\perp \sum_{i} \left(\chi^{\perp\dag}_{iz}\left\langle\chi^\perp_{z}\right\rangle + \mathrm{h.c.} - \left\langle\chi^{\perp\dag}_{z}\right\rangle\left\langle\chi^\perp_{z}\right\rangle\right)\\
        &-\frac{3}{8} J_\perp \sum_{i} \left(\Delta^{\perp\dag}_{iz}\left\langle\Delta^\perp_{z}\right\rangle + \mathrm{h.c.} - \left\langle\Delta^{\perp\dag}_{z}\right\rangle\left\langle\Delta^\perp_{z}\right\rangle\right)\\
        &- \frac{3}{8} \tilde{J}_\perp \sum_{i} \left(\Delta^{\perp\dag}_{ix}\langle\Delta^\perp_{x}\rangle + \mathrm{h.c.} - \langle\Delta^{\perp\dag}_{x}\rangle\langle\Delta^\perp_{x}\rangle\right) \\
        &+ \sum_{i\mu\alpha\sigma}\epsilon_\alpha n_{i\mu \alpha\sigma} + \frac{\varepsilon}{2} \sum_{i\alpha\sigma} n_{it\alpha\sigma} - \frac{\varepsilon}{2} \sum_{i\alpha\sigma}n_{ib\alpha\sigma}.
    \end{aligned}
\end{equation}

where $\delta_{\mu\alpha}=\left\langle{}b_{i\mu\alpha}b^{\dag}_{j\mu\alpha}\right\rangle$ since holon condense at zero temperature. Under the electric field, we have $\delta_{bz}=0$ and $\delta_{tz}$, $\delta_{tx}$ and $\delta_{bx}$ are solved in a self-consistent manner by adjustment to onsite energies $\epsilon_\alpha$ (See SI \cite{SM} for more details). The mean-field order parameters are represented by
\begin{equation}
    \begin{aligned}
        &\chi_{ij,\mu x} = \sum_{\sigma} f_{i\mu x\sigma}^\dag f_{j\mu x\sigma},\\ 
        &\chi^{\perp\dag}_{iz} = \sum_\sigma f^\dag_{izt\sigma}f_{izb\sigma},\\
        &\Delta^\dag_{ij,\mu\alpha} = f^\dag_{i\mu\alpha\uparrow}f^\dag_{j\mu\alpha\downarrow}-f^\dag_{i\mu\alpha\downarrow}f^\dag_{j\mu\alpha\uparrow},\\
        &\Delta^{\perp\dag}_{i\alpha} = f^\dag_{it\alpha\uparrow}f^\dag_{ib\alpha\downarrow}-f^\dag_{ib\alpha\downarrow}f^\dag_{it\alpha\uparrow},
    \end{aligned}
\end{equation}
and
\begin{equation}
    \begin{aligned}
        &\chi_{\mu x} = \frac{1}{2N}\sum_{\langle i,j\rangle}\chi_{ij,\mu x},\ \chi^\perp_z = \frac{1}{N}\sum_i\chi^\perp_{iz},\\
        &\Delta^{\mathbf{x}}_{\mu x} = \frac{1}{2N}\sum_{\langle i,j\rangle}\Delta_{ij,\mu x},\ \Delta^\perp_\alpha = \frac{1}{N}\sum_i\Delta^\perp_{i\alpha}.
    \end{aligned}
\end{equation}

Notably, the spin-exchange $\tilde{J}_\perp$ of the Hamiltonian in Eq. (\ref{eq_H2}) doesn't produce a hopping term $\chi_{x}^\perp$ in Eq. (\ref{eq_MF2}), which is the feature of such a bilayer system. Without interlayer hopping, a small interlayer spin-exchange $J_\perp$ leads to $\langle\chi^\perp\rangle\approx0$.

Consequently, the $3d_{z^2}$ orbital only participate in the interlayer pairing. However, this pairing is not SC as the corresponding SC order parameter goes to zero in the SBMF theory due to $\delta_{bz}=0$. The SC is carried by the $3d_{x^2-y^2}$ orbitals, which can form both intralayer and interlayer pairing. The superconducting $T_c$ scales with the ground state gap amplitude of the $3d_{x^2-y^2}$ orbitals via the BCS relation exhibited in Fig.~3(b).

The expectation value of the mean-field order parameters are obtained by numerically solving the following self-consistent equations
\begin{equation}
\begin{aligned}
    &\delta_{\mu\alpha} = 1-\frac{1}{N}\sum_k \left(\langle f^\dag_{k\mu\alpha\uparrow}f_{k\mu\alpha\uparrow}\rangle + \langle f^\dag_{-k\mu\alpha\downarrow}f_{-k\mu\alpha\downarrow}\rangle\right),\\
    &\delta_{tz} = 0,\ \sum_{\mu\alpha}\delta_{\mu\alpha} = 1,\\
    &\langle\chi_{\mu x}\rangle = \frac{1}{N}\sum_k \epsilon(\mathbf{k})\left(\langle f^\dag_{k\mu x\uparrow}f_{k\mu x\uparrow}\rangle + \langle f^\dag_{-k\mu x\downarrow}f_{-k\mu x\downarrow}\rangle\right),\\
    &\langle\chi^\perp_z\rangle = \frac{1}{N}\sum_k\left(\langle f^\dag_{ktz\uparrow}f_{kbz\uparrow}\rangle + \langle f^\dag_{-ktz\downarrow}f_{-kbz\downarrow}\rangle\right),\\
    &\langle\Delta^{\mathbf{x}}_{\mu x}\rangle^* = \frac{1}{N}\sum_{k}2\cos{(k_x)}\langle f^\dag_{k\mu x\uparrow}f^\dag_{-k\mu x\downarrow}\rangle,\\
    &\langle\Delta^\perp_\alpha\rangle^* = \frac{2}{N}\sum_k \langle f^\dag_{kt\alpha\uparrow}f^\dag_{-kb\alpha\downarrow}\rangle,
\end{aligned}
\end{equation}
where $\epsilon(\mathbf{k}) = \frac{\cos{(k_x)}+\cos{(k_y)}}{2}$.


~~~

\section{Data availability}
The data generated in this study have been deposited in the Zenodo database at https://zenodo.org/records/17691236. 

\section{Code availability}
The code that supports the plots within this paper is available from the corresponding author upon request.

~~~~~~~~~~~~~


~~~~~~~~~~~~~

\section{\bf Acknowledgement}
We are grateful to the stimulating discussions with Chen Lu. 
F. Y. and C. W. is supported by the National Natural Science Foundation of China (NSFC) under the Grant No. 12234016. 
F. Y. is also supported by the CAS Superconducting Research Project under Grant No. [SCZX-0101] and the NSFC under the Grant No. 12074031. 
C. W. is also supported by the NSFC under the Grant No. 12174317. 
D. X. Y. is supported by NSFC-12494591, NSFC-92165204, NSFC-92565303, NKRDPC-2022YFA1402802, Research Center for Magnetoelectric Physics of Guangdong Province (2024B0303390001), and Guangdong Provincial Quantum Science Strategic Initiative (GDZX2401010). 
C. W. is also supported by the New Cornerstone Science Foundation. 

\section{Author contributions}
F. Yang proposed the main idea and supervised the study.
Z.-Y. Shao performed the SBMF study.
J.-H. Ji performed the DMRG study. D.-X. Yao and C. Wu helped shape the main idea. F. Yang, Z.-Y. Shao and J.-H. Ji wrote the paper.

\section{Additional Information}

\paragraph{Competing Interests}
The authors declare no competing interests.

~~~~~

\newpage

\onecolumngrid

\section*{Supplementary Information}

\onecolumngrid
\renewcommand{\theequation}{S\arabic{equation}}
\renewcommand{\thefigure}{A\arabic{figure}}
\setcounter{equation}{0}
\setcounter{figure}{0}

\section{A. Slave-boson mean-field treatment of the one-orbital model}
\label{appendix_a}
In the one-orbital model, we begin with the Hamiltonian
\begin{equation}\label{ap_eq_H1}
    H=-t_{\parallel} \sum_{\left\langle{}i,j\right\rangle,\mu,\sigma} \hat{P} \left( c^{\dag}_{i\mu\sigma}c_{j\mu\sigma} + \mathrm{h.c.} \right) \hat{P} + \sum_{i,\mu} \epsilon_{\mu} n_{i\mu} + J_{\parallel} \sum_{\left\langle{}i,j\right\rangle,\mu} \mathbf{S}_{i\mu} \cdot \mathbf{S}_{j\mu} +  \tilde{J}_{\perp} \sum_{i} \mathbf{S}_{it} \cdot \mathbf{S}_{ib},
 \end{equation}
where $c_{i\mu\sigma}^\dag$ creates a electron in the $d_{x^2-y^2}$ orbital with the spin $\sigma = \{\uparrow,\downarrow\}$ at the lattice site $i$ in the layer $\mu=\{t,b\}$. $n_{i\mu} = \sum_{\sigma}c_{i\mu\sigma}^\dag c_{i\mu\sigma}$ is the particle number operator. $\mathbf{S}{}_{i\mu} = \frac{1}{2} c^{\dag}_{i\mu\sigma} \left[\mathbf{\sigma}\right]_{\sigma\sigma^\prime} c_{i\mu\sigma^\prime}$ is the spin operator with Pauli matrix $\mathbf{\sigma}=\left(\sigma_x,\sigma_y,\sigma_z\right)$. $\hat{P}$ is a projection operator projecting out the double occupancy of all site. Here we set $t_\parallel=1$ as the unit, and the intralayer and interlayer spin exchange are given by $J_\parallel=0.4t_\parallel$ and $\tilde{J}_{\perp}\approx \alpha J_{\perp} (1-\delta_{tz})$ with $J_{\perp}=1.3J_{\parallel}$ and $\alpha=1$, where $\delta_{tz}$ denotes the hole density of the top-3$d_{z^2}$ orbital. The particle number of the bottom-layer $d_{z^2}$ orbital is near half-filling without external electric field, thus in the strong-coupling limit, the bottom-layer $d_{z^2}$ orbital approaches half-filling and becomes incapable of accommodating additional electrons, even under a small perpendicular electric field. Therefore we fix the particle number of the bottom-layer $d_{z^2}$ orbital $n_{bz}=1$. Considering that the electrons flow from the top layer only to the bottom layer $d_{x^2-y^2}$ orbital, we express the particle number of the top-layer $d_{z^2}$ orbital, the top-layer $d_{x^2-y^2}$ orbital and the bottom layer $d_{x^2-y^2}$ orbital as $n_{tz}=1-\delta$, $n_{tx}=0.5-\eta_1$ and $n_{bx}=0.5+\delta+\eta_1=0.5+\eta_2$ respectively. In this simplified single-orbital study, we cannot determine the concrete relation between $\varepsilon$ and $\delta$ or $\eta_{1,2}$. But clearly, $\delta$ and $\eta_1$ enhance with the enhancement of $\varepsilon$. Thus we set $\delta$ and $\eta_1$ as certain numbers and $\epsilon_\mu$ corresponding to chemical potential is consistently derived from $\delta$, then $\varepsilon=\epsilon_t-\epsilon_b$.

To get the knowledge of $\eta_1/\delta$ in different perpendicular electric field, we solve the tight-binding (TB) Hamiltonian 
with TB parameters reported in Ref. \cite{Daoxin_Yao2025}, i.e. $t_{\parallel}=0.445\text{ eV}$, $t_{xz}=0.221\text{ eV}$, $t_{\perp}=0.503\text{ eV}$ and $\epsilon_x-\epsilon_z=0.367\text{ eV}$, 
and an additional external electric field term $(\varepsilon/2)\sum_{i,\alpha,\sigma}(n_{it\alpha\sigma}-n_{ib\alpha\sigma})$. We plot the $(\delta, \eta_1)$ points (Fig. \ref{fig_ParticleNumber_TB}), finding that $\eta_1/\delta\approx1/2$. Thus we assume $\eta_1/\delta=1/2$ (i.e. $(n_{tx}-0.5)/(n_{bx}-0.5)=-\eta_1/\eta_2=-1/3$) in the one-orbital model study.

\begin{figure}[htbp]
    \centering
    \includegraphics[width=0.36\linewidth]{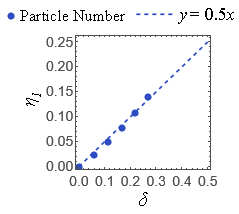}
    \caption{The change of electron number $\delta$ and $\eta_1$ under different electric field. The results are represented by the dots $(\delta, \eta_1)$. All of the dots are near the $y=0.5x$ line (dashed line), indicating that $\eta_1 \approx 0.5\delta$.} \label{fig_ParticleNumber_TB}
\end{figure}

We solve the Hamiltonian in Eq. (\ref{ap_eq_H1}) by slave-boson mean-field (SBMF) theory~\cite{kotliar1988}. The electron operator ($c$) is decomposed into the product of a fermonic spinon operator ($f$) and a bosonic holon operator ($b$), i.e. $c^{\dag}_{i\mu\sigma}=f^{\dag}_{i\mu\sigma}b_{i\mu}$. 
Then the electron pairing operator is expressed as 
\begin{equation}
    \Delta_e\equiv{}cc=b^{\dagger}b^{\dagger}ff, 
\end{equation}
and the expectation value is the product of the expectation value of the holon operators and the one of the spinon pairing operator (defined as $\Delta\equiv{}ff$), 
\begin{equation}
    \left\langle\Delta_e\right\rangle = \left\langle{}b^{\dagger}b^{\dagger} \right\rangle\left\langle{}ff\right\rangle. 
\end{equation}
When the temperature is below the holon condensation temperature $T_{\mathrm{BEC}}$, holons condense and $\left\langle{}b\right\rangle=\left\langle{}b^{\dagger}\right\rangle=\sqrt{\delta}$, where $\delta$ is the hole density, while $\left\langle{}b\right\rangle=\left\langle{}b^{\dagger}\right\rangle=0$ when the temperature is above $T_{\mathrm{BEC}}$. 
\begin{equation}
    \left\langle{}b\right\rangle = \left\langle{}b^{\dagger}\right\rangle = 
    \begin{cases}
        \sqrt{\delta}, & T<T_{\mathrm{BEC}} \\
        0, & T>T_{\mathrm{BEC}}
    \end{cases}
\end{equation}
The spinon pairing operator $\Delta\equiv{}ff$ has a non-zero expectation value when the temperature is below the spinon pairing temperature $T_{\mathrm{pair}}$ while the expectation value is zero when the temperature is above $T_{\mathrm{pair}}$. 
\begin{equation}
    \left\langle\Delta\right\rangle = \left\langle{}ff\right\rangle 
    \begin{cases}
        >0, & T<T_{\mathrm{pair}} \\
        =0, & T>T_{\mathrm{pair}}
    \end{cases}
\end{equation}
Thus, the electron pairing operator $\Delta_e$ has a non-zero expectation value if and only if both $\left\langle{}b^{\dagger}b^{\dagger}\right\rangle$ and $\left\langle\Delta\right\rangle$ are non-zero, i.e. if and only if the temperature is below both $T_{\mathrm{BEC}}$ and $T_{\mathrm{pair}}$. Therefore, the critical temperature of superconductivity (SC), $T_c$, is the lower one between $T_{\mathrm{BEC}}$ and $T_{\mathrm{pair}}$, 
\begin{equation}
    T_c=\mathrm{min}(T_{\mathrm{BEC}},T_{\mathrm{pair}}).
\end{equation}
The method for determining $T_{\mathrm{pair}}$ and $T_{\mathrm{BEC}}$ is given below. For the doping level considered in this work, $T_{\mathrm{BEC}}\sim10^{3}~\mathrm{K}\gg{}T_{\mathrm{pair}}$ (see the main text), therefore $T_c=T_{\mathrm{pair}}$.

The mean-field Hamiltonian of spinon and holon can be expressed as
\begin{equation}\label{eq_spinon}
\begin{aligned}
    H_\mathrm{spinon} = & -t_\parallel \sum_{\left\langle i,j\right\rangle,\mu,\sigma} \left( \left\langle b_{i\mu}b^{\dag}_{j\mu}\right\rangle{} f^{\dag}_{i\mu\sigma}f_{j\mu\sigma} + \mathrm{h.c.} \right) + \sum_{i,\mu,\sigma} \epsilon_{\mu} f^{\dag}_{i\mu\sigma}f_{i\mu\sigma}\\
    & - \frac{3}{8} J_\parallel \sum_{\left\langle i,j\right\rangle,\mu} \left(  \left\langle\chi^{\dag}_{\mu}\right\rangle \chi_{ij,\mu} + \mathrm{h.c.}  + \left\langle\Delta^{\dag}_{\mu}\right\rangle \Delta_{ij,\mu} + \mathrm{h.c.}  - \left\langle\chi^{\dag}_{\mu}\right\rangle \left\langle\chi_{\mu}\right\rangle - \left\langle\Delta^{\dag}_{\mu}\right\rangle \left\langle\Delta_{\mu}\right\rangle \right) \\
    & - \frac{3}{8} \left(1-\delta\right) J_{\perp} \sum_{i} \left(  \left\langle\chi^{\perp\dag}\right\rangle \chi^\perp_{i} + \mathrm{h.c.} + \left\langle\Delta^{\perp\dag}\right\rangle \Delta^\perp_{i} + \mathrm{h.c.} - \left\langle\chi^{\perp\dag}\right\rangle \left\langle\chi^\perp\right\rangle - \left\langle\Delta^{\perp\dag}\right\rangle \left\langle\Delta^\perp\right\rangle \right),
\end{aligned}
\end{equation}
and
\begin{equation}\label{eq_holon}
    H_{\mathrm{holon}} = -t_\parallel \sum_{\left\langle i,j\right\rangle,\mu} \left( \left\langle\chi_{ij,\mu}\right\rangle b^{\dag}_{i\mu}b_{j\mu} + \mathrm{h.c.} \right),
\end{equation}
where $\left\langle{}b_{i\mu}\right\rangle=\left\langle{}b^{\dag}_{i\mu}\right\rangle=\sqrt{\delta_{\mu}}$ is sustained below the holon condensation temperature, and the bonding and pairing order parameters are defined as
\begin{equation}\label{eq_op1}
\begin{aligned}
    &\chi^{\dag}_{ij,\mu} = \sum_{\sigma} f^{\dag}_{i\mu\sigma}f_{j\mu\sigma},\ \chi^{\dag}_{\mu} = \frac{1}{2N}\sum_{\langle i,j\rangle}\chi^{\dag}_{ij,\mu},\\
    &\chi^{\perp\dag}_{i} = \sum_{\sigma} f^{\dag}_{it\sigma}f_{ib\sigma},\ \chi^{\perp\dag} = \frac{1}{N}\sum_i\chi^{\perp\dag}_{i},\\
    &\Delta^{\dag}_{ij,\mu} = f^{\dag}_{i\mu\uparrow}f^{\dag}_{j\mu\downarrow} - f^{\dag}_{i\mu\downarrow}f^{\dag}_{j\mu\uparrow},\ 
    \Delta^{\mathbf{x}(\mathbf{y})}_{\mu} = \frac{1}{2N}\sum_{\mathbf{R}_i-\mathbf{R}_j=\pm\mathbf{x}(\mathbf{y})}\Delta_{ij,\mu},\\
    &\Delta^{\perp\dag}_{i} = f^{\dag}_{it\uparrow}f^{\dag}_{ib\downarrow} - f^{\dag}_{it\downarrow}f^{\dag}_{ib\uparrow},\ \Delta^\perp = \frac{1}{N}\sum_i\Delta^\perp_{i}.
\end{aligned}
\end{equation}
Here, $N$ is site number and $\mathbf{x}/\mathbf{y}$ is the basis vector along the $+x/y$ direction. 

In the mean-field approach, we constrain the particle number and obtain the expectation value of the mean-field order parameters 
by solving the self-consistent equations derived from Eq. (\ref{eq_spinon}). 
\begin{equation}\label{ap_SC1}
\begin{aligned}
    &\delta_{\mu} = 1-\frac{1}{N}\sum_k \left(\langle f^\dag_{k\mu\uparrow}f_{k\mu\uparrow}\rangle + \langle f^\dag_{-k\mu\downarrow}f_{-k\mu\downarrow}\rangle\right),\ \delta_t=0.5+\eta_1,\ \delta_b=0.5-\eta_2,\\
    &\langle\chi_{\mu}\rangle = \frac{1}{2N}\sum_{\langle i,j\rangle}\langle\chi_{ij,\mu}\rangle = \frac{1}{N}\sum_k \frac{\cos{(k_x)}+\cos{(k_y)}}{2}\left(\langle f^\dag_{k\mu\uparrow}f_{k\mu\uparrow}\rangle + \langle f^\dag_{-k\mu\downarrow}f_{-k\mu\downarrow}\rangle\right),\\
    &\langle\chi^\perp\rangle = \frac{1}{N}\sum_i\langle\chi^\perp_{i}\rangle = \frac{1}{N}\sum_k\left(\langle f^\dag_{kt\uparrow}f_{kb\uparrow}\rangle + \langle f^\dag_{-kt\downarrow}f_{-kb\downarrow}\rangle\right),\\
    &\langle\Delta^{\mathbf{x}(\mathbf{y})}_{\mu}\rangle^* = \frac{1}{2N}\sum_{\mathbf{R}_i-\mathbf{R}_j=\pm\mathbf{x}(\mathbf{y})}\langle\Delta_{ij,\mu}\rangle^* = \frac{1}{N}\sum_{k}2\cos{(k_{x(y)})}\langle f^\dag_{k\mu\uparrow}f^\dag_{-k\mu\downarrow}\rangle,\\
    &\langle\Delta^\perp\rangle^* = \frac{1}{N}\sum_i\langle\Delta^\perp_{i}\rangle^* = \frac{2}{N}\sum_k \langle f^\dag_{kt\uparrow}f^\dag_{-kb\downarrow}\rangle.
\end{aligned}
\end{equation}
Here, $f_{\mathbf{k}\mu\sigma}=\frac{1}{\sqrt{N}}\sum_{i}f_{i\mu\sigma}\mathrm{e}^{-\mathrm{i}\mathbf{k}\cdot\mathbf{R}_i}$, $k_{x/y}$ is the $x/y$-component of $\mathbf{k}$, and $\left\langle\Delta_{\mu}^{\mathbf{x}/\mathbf{y}}\right\rangle$ is the gap of the nearest neighbor bond along the $x/y$ direction. For $s$-wave pairing, $\left\langle\Delta_{\mu}^{\mathbf{y}}\right\rangle=\left\langle\Delta_{\mu}^{\mathbf{x}}\right\rangle$. For $d$-wave pairing, $\left\langle\Delta_{\mu}^{\mathbf{y}}\right\rangle=-\left\langle\Delta_{\mu}^{\mathbf{x}}\right\rangle$. 

In the self consistent equations, the expectation values of the operators are obtained as follows. The matrix form of the spinon Hamiltonian is 
\begin{equation}
    H_{\mathrm{spinon}} = \sum_{k}
    \left(
    \begin{array}{cccc}
        f^{\dagger}_{kt\uparrow} & f^{\dagger}_{kb\uparrow} & f_{-kt\downarrow} & f_{-kb\downarrow}
    \end{array}
    \right)
    \left(
    \begin{array}{cc}
        H_{\chi}(\mathbf{k}) & H_{\Delta}(\mathbf{k}) \\
        H_{\Delta}^{\dagger}(\mathbf{k}) & -H_{\chi}^T(\mathbf{k})
    \end{array}
    \right)
    \left(
    \begin{array}{c}
        f_{kt\uparrow} \\ f_{kb\downarrow} \\ f^{\dagger}_{-kt\downarrow} \\ f^{\dagger}_{-kb\downarrow}
    \end{array}
    \right), 
\end{equation}
where 
\begin{equation}
    \begin{aligned}
        & H_{\chi}(\mathbf{k}) = 
        \left(
        \begin{array}{cc}
            -2 \left( t_{\parallel}\delta_t + \frac{3}{8}J_{\parallel}\left\langle\chi_t\right\rangle \right) \left( \cos(k_x) + \cos(k_y) \right) + \epsilon_t & -\frac{3}{8}\left(1-\delta\right)J_{\perp}\left\langle\chi^{\perp}\right\rangle \\
            -\frac{3}{8}\left(1-\delta\right)J_{\perp}\left\langle\chi^{\perp}\right\rangle & -2 \left( t_{\parallel}\delta_b + \frac{3}{8}J_{\parallel}\left\langle\chi_b\right\rangle \right) \left( \cos(k_x) + \cos(k_y) \right) + \epsilon_b
        \end{array}
        \right), \\
        & H_{\Delta}(\mathbf{k}) = 
        \left(
        \begin{array}{cc}
            -\frac{3}{4}J_{\parallel} \left( \left\langle\Delta_t^{\mathbf{x}}\right\rangle\cos(k_x) + \left\langle\Delta_t^{\mathbf{y}}\right\rangle\cos(k_y) \right) & -\frac{3}{8}\left(1-\delta\right)J_{\perp} \left\langle\Delta^{\perp}\right\rangle \\
            -\frac{3}{8}\left(1-\delta\right)J_{\perp} \left\langle\Delta^{\perp}\right\rangle & -\frac{3}{4}J_{\parallel} \left( \left\langle\Delta_b^{\mathbf{x}}\right\rangle\cos(k_x) + \left\langle\Delta_b^{\mathbf{y}}\right\rangle\cos(k_y) \right)
        \end{array}
        \right). 
    \end{aligned}
\end{equation}

$H_{\mathrm{spinon}}$ can be diagonalized by a Bogoliubov transformation. 
\begin{equation}
    H_{\mathrm{spinon}} = \sum_{k} 
    \left(
    \begin{array}{cccc}
        \gamma^{\dagger}_{k1} & \gamma^{\dagger}_{k2} & \gamma^{\dagger}_{k3} & \gamma^{\dagger}_{k4}
    \end{array}
    \right) \ 
    \mathrm{diag}(E_{k1},E_{k2},E_{k3},E_{k4}) \ 
    \left(
    \begin{array}{c}
        \gamma_{k1} \\ \gamma_{k2} \\ \gamma_{k3} \\ \gamma_{k4}
    \end{array}
    \right), 
\end{equation}
where
\begin{equation}
    \xi^{\dagger}(\mathbf{k}) 
    \left(
    \begin{array}{cc}
        H_{\chi}(\mathbf{k}) & H_{\Delta}(\mathbf{k}) \\
        H_{\Delta}^{\dagger}(\mathbf{k}) & -H_{\chi}^T(\mathbf{k})
    \end{array}
    \right)
    \xi(\mathbf{k}) 
    = \mathrm{diag}(E_{k1},E_{k2},E_{k3},E_{k4}), \ 
    \left(
    \begin{array}{c}
        f_{kt\uparrow} \\ f_{kb\uparrow} \\ f^{\dagger}_{-kt\downarrow} \\ f^{\dagger}_{-kb\downarrow}
    \end{array}
    \right) 
    = \xi(\mathbf{k}) 
    \left(
    \begin{array}{c}
        \gamma_{k1} \\ \gamma_{k2} \\ \gamma_{k3} \\ \gamma_{k4}
    \end{array}
    \right). 
\end{equation}
Here, $\gamma$ is fermionic operator, whose expectation value is 
\begin{equation}
    \left\langle\gamma^{\dagger}_{km}\gamma_{k^{\prime}n}\right\rangle = \delta_{kk^{\prime}}\delta_{mn}n_F(E_{km}), \ 
    \left\langle\gamma_{km}\gamma_{k^{\prime}n}\right\rangle = \left\langle\gamma^{\dagger}_{km}\gamma^{\dagger}_{k^{\prime}n}\right\rangle = 0, 
\end{equation}
where $m=1,2,3,4$ and $n_F(E_{km})=1/\left(\mathrm{e}^{E_{km}/k_BT}+1\right)$ is Fermi distribution function with $T$ representing the temperature. Then the order parameter operators are expressed as 
\begin{equation} \label{ap_SC1_T}
    \begin{aligned}
        & \delta_{t(b)} = 1 - \frac{1}{N} \sum_{k} \sum_{m=1,2,3,4} \left[ \xi^{\ast}_{1(2),m}(\mathbf{k})\xi_{1(2),m}(\mathbf{k})n_F(E_{km}) + \xi^{\ast}_{3(4),m}(\mathbf{k})\xi_{3(4),m}(\mathbf{k})\left(1-n_F(E_{km})\right) \right], \\
        & \left\langle\chi_{t(b)}\right\rangle = \frac{1}{N} \sum_{k} \frac{\cos(k_x)+\cos(k_y)}{2} \sum_{m=1,2,3,4} \left[ \xi^{\ast}_{1(2),m}(\mathbf{k})\xi_{1(2),m}(\mathbf{k})n_F(E_{km}) + \xi^{\ast}_{3(4),m}(\mathbf{k})\xi_{3(4),m}(\mathbf{k})\left(1-n_F(E_{km})\right) \right], \\
        & \left\langle\chi^{\perp}\right\rangle = \frac{1}{N} \sum_{k} \sum_{m=1,2,3,4} \left[ \xi^{\ast}_{1,m}(\mathbf{k})\xi_{2,m}(\mathbf{k})n_F(E_{km}) + \xi^{\ast}_{4,m}(\mathbf{k})\xi_{3,m}(\mathbf{k})\left(1-n_F(E_{km})\right) \right], \\
        & \left\langle\Delta_{t(b)}^{\mathbf{x}/\mathbf{y}}\right\rangle^{\ast} = \frac{1}{N} \sum_{k} 2\cos(k_{x/y}) \sum_{m=1,2,3,4} \xi^{\ast}_{1(2),m}(\mathbf{k})\xi_{3(4),m}(\mathbf{k})n_F(E_{km}), \\
        & \left\langle\Delta^{\perp}\right\rangle = \frac{2}{N} \sum_{k} \sum_{m=1,2,3,4} \xi^{\ast}_{1,m}(\mathbf{k})\xi_{4,m}(\mathbf{k})n_F(E_{km}),
    \end{aligned}
\end{equation}
where $\xi_{a,m}$ represents the matrix element of $\xi$ at the $a$-th row and the $m$-th column. 

By solving Eq. (\ref{ap_SC1_T}) at zero temperature, we obtain the zero-temperature spinon pairing gap.
From Eq. (\ref{ap_SC1_T}), we can also determine the spinon-pairing temperature $T_\text{pair}$ by solving for the critical condition $\Delta=0$ at finite temperature.
The self-consistent equations are solved by iteration, which is equivalent to finding the minimum value of energy. When solving the self-consistent equations, we assume the pairing symmetry and obtain the energy of the corresponding state. Then the pairing symmetry is determined by the pairing state with the lowest energy. Here we show the energy per site of the $s$-wave pairing state and the $d$-wave pairing state relative to the non-superconducting normal state, $(E_s-E_n)/N$ and $(E_d-E_n)/N$, and their difference $(E_s/N)-(E_d/N)$ as functions of the bottom layer $d_{x^2-y^2}$ particle number $n_{bx}$ (Fig. \ref{fig_SBMF_1orb_EsEd}). We also consider the $s+\mathrm{i}d$-wave pairing, whose expression is actually $\alpha{}s+\mathrm{i}\beta{}d$, where $\alpha$ and $\beta$ are variational parameters determined by energy minimization. When one of $\alpha$ and $\beta$ is zero, the mixed state decays to the $s$- or $d$- wave state. For most $n_{bx}$, we find that one of $\alpha$ and $\beta$ is zero. 

\begin{figure}[htbp]
    \centering
    \includegraphics[width=0.7\linewidth]{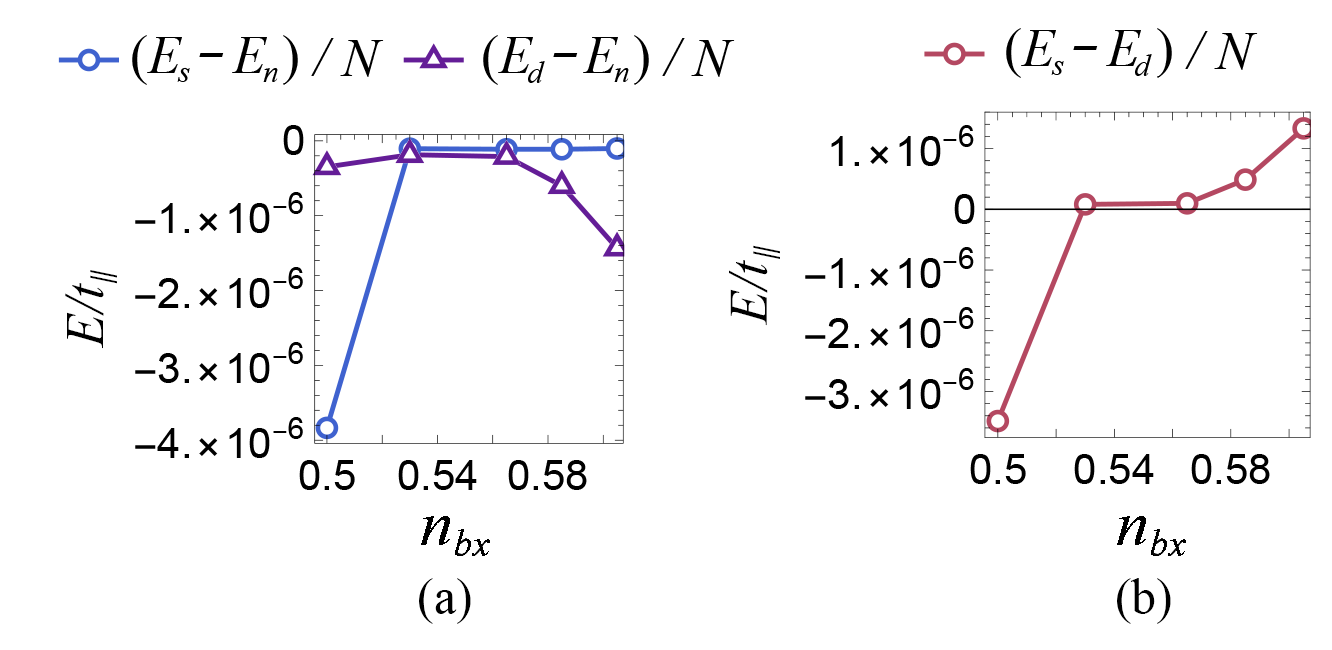}
    \caption{(a) Energy of the $s$-wave pairing state relative to that of the non-superconducting normal state  $(E_s-E_n)/N$ and that of the $d$-wave pairing state relative to that of the normal state $(E_d-E_n)/N$, and (b) their difference $(E_s/N)-(E_d/N)$ as functions of the bottom layer $d_{x^2-y^2}$ particle number $n_{bx}$ obtained by the SBMF study for the one orbital model. } \label{fig_SBMF_1orb_EsEd}
\end{figure}

To characterize the pairing type of the system, the pairing gap amplitude $\tilde{\Delta}$ is defined as the maximal value of the pairing gaps. For the interlayer $s$-wave pairing, $\tilde{\Delta}=-\frac{3}{8}\tilde{J}_{\perp}\left\langle\Delta^{\perp}\right\rangle$. For the $d$-wave pairing, $\tilde{\Delta}=-\frac{3}{2}J_{\parallel}\left\langle\Delta^{\mathbf{x}}_{b}\right\rangle$.
  
The holon condensation temperature can be calculated according to the Berezinskii-Kosterlitz-Thouless (BKT) transition theory. The holon operator can be written as $b^{\dag}_{i\mu}=\sqrt{\delta_{\mu}}\mathrm{e}^{\mathrm{i}\theta_{\mu}(i)}$, whose phase fluctuations lead to the BKT transition. Then the holon Hamiltonian can be written as a $XY$-model-like form
\begin{equation}
    H_{\mathrm{holon}} = - 2 t_{\parallel} \sum_{\left\langle{}i,j\right\rangle,\mu} \left\langle\chi_{\mu}\right\rangle \delta_{\mu} \cos\left(\theta_{\mu}(i)-\theta_{\mu}(j)\right).
\end{equation}
We transform the model to a continuous model
\begin{equation}
    H_{\mathrm{holon}} \sim \frac{1}{2} \rho \int \mathrm{d}^2{\mathbf{r}} \left|\nabla\theta\right|^2,
\end{equation}
where 
\begin{equation}
    \rho = \sum_{\left|\mathbf{l}\right|=1} \sum_{\mu} t_{\parallel} \left\langle\chi_{\mu}\right\rangle \delta_{\mu} l^2
\end{equation}
is the superfluid stiffness. Then we can get $T_{\mathrm{BEC}}$ from the relationship $T_{\mathrm{BEC}}=(\pi/2)\rho$.

Considering that the real $\eta_1/\delta$ may be different from the TB results, we also study the $\eta_1/\delta=0/1$ and $\eta_1/\delta=-1/2$ situations (i.e. $(n_{tx}-0.5)/(n_{bx}-0.5)=0/1$ and $(n_{tx}-0.5)/(n_{bx}-0.5)=1/2$). We find that the results of different $(n_{tx}-0.5)/(n_{bx}-0.5)$ are similar to each other (shown in Fig. \ref{fig_m13_01_12} (a)-(c)). Additionally, the $d$-wave pairing gap $\tilde{\Delta}^{d}$ is only determined by the particle number $n_{bx}$ but is not related to the particle number ratio $(n_{tx}-0.5)/(n_{bx}-0.5)$ (shown in Fig. \ref{fig_m13_01_12} (d)).

\begin{figure}[htbp]
    \centering
    \includegraphics[width=0.96\linewidth]{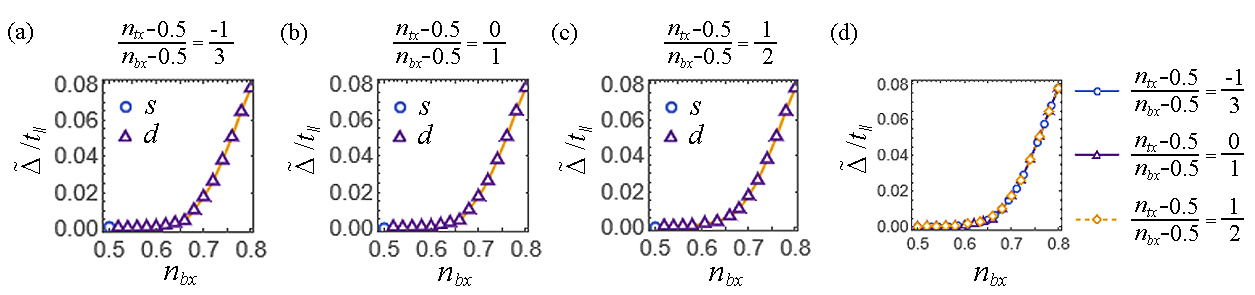}
    \caption{(a)-(c) The pairing amplitude $\tilde{\Delta}$ (in unit of $t_{\parallel}$) as function of the bottom-layer particle number $n_{bx}$ controlled by the imposed electric field. Different pairing symmetry is marked by different colors. (d) The pairing gap $\tilde{\Delta}$ of different particle number ratios as function of particle number $n_{bx}$ (represented by different lines). The lines almost coincide with each other especially at the $d$-wave pairing regime, indicating that the $d$-wave pairing gap is only determined by the particle number $n_{bx}$.} \label{fig_m13_01_12}
\end{figure}

We also study how different $J_{\perp}$ affects the result. The relationship between the pairing gap and $n_{bx}$ is calculated when $J_{\perp}=1.8J_{\parallel}$ as well. The results are shown in Fig. \ref{fig_1orb_Jperp}. The results of different $J_{\perp}$ are similar to each other and they almost coincide with each other especially at the $d$-wave pairing regime when plotted together. This indicates that the $d$-wave pairing gap is not related to $J_{\perp}$, neither. 

\begin{figure}[htbp]
    \centering
    \includegraphics[width=0.8\linewidth]{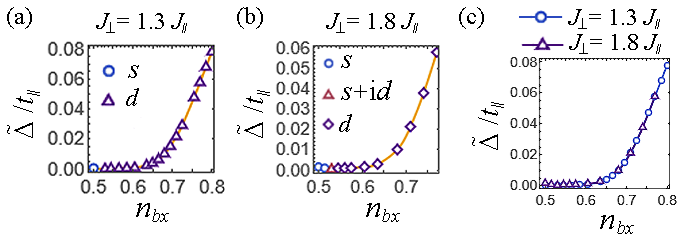}
    \caption{(a)-(b) The pairing amplitude $\tilde{\Delta}$ (in unit of $t_{\parallel}$) as function of the bottom-layer particle number $n_{bx}$ controlled by the imposed electric field. Different pairing symmetry is marked by different colors. (c) The pairing gap $\tilde{\Delta}$ of $J_{\perp}$ as function of particle number $n_{bx}$ (represented by different lines). The lines almost coincide with each other especially at the $d$-wave pairing regime, indicating that the $d$-wave pairing gap is only determined by the particle number $n_{bx}$.} \label{fig_1orb_Jperp}
\end{figure}

\section{B. Slave-boson mean-field treatment of the two-orbital model} 
\label{appendix_b}
In the two-orbital model, the Hamiltonian takes the form
\begin{equation}\label{ap_eq_H2}
\begin{aligned}
    H = &- t_\parallel\sum_{\langle i,j\rangle,\mu} \hat{P} \left(c^\dag_{i\mu x\sigma}c_{j\mu x\sigma} + \mathrm{h.c.}\right) \hat{P} - t_\perp\sum_i \hat{P} \left(c^\dag_{itz\sigma}c_{ibz\sigma} + \mathrm{h.c.}\right) \hat{P} - t_{xz} \sum_{\langle i,j\rangle\mu} \hat{P} \left(c^\dag_{i\mu x\sigma}c_{j\mu z\sigma}+(z\leftrightarrow x)+\mathrm{h.c.}\right) \hat{P} \\
    &+ J_\parallel\sum_{\langle i,j\rangle\mu}\mathbf{S}_{i\mu x}\cdot\mathbf{S}_{j\mu x} + J_\perp\sum_{i}\mathbf{S}_{itz}\cdot\mathbf{S}_{ibz} + \tilde{J}_\perp\sum_{i}\mathbf{S}_{it x}\cdot\mathbf{S}_{ib x}+\epsilon_z\sum_{i\mu\sigma}n_{i\mu z\sigma} + \epsilon_x\sum_{i\mu\sigma}n_{i\mu x\sigma} \\
    &+\frac{\varepsilon}{2} \sum_{i\alpha\sigma} n_{it\alpha\sigma} - \frac{\varepsilon}{2} \sum_{i\alpha\sigma}n_{ib\alpha\sigma},
\end{aligned}
\end{equation}
where $c_{i\mu z\sigma}^\dag / c_{i\mu x\sigma}^\dag$ creates a $d_{z^2} / d_{x^2-y^2}$-orbital electron with the spin $\sigma = \{\uparrow,\downarrow\}$ at the lattice site $i$ in the layer $\mu=\{t,b\}$. $n_{i\mu\alpha\sigma} = c_{i\mu\alpha\sigma}^\dag c_{i\mu\alpha\sigma}$ is the particle number operator for the two $E_g$ orbitals with $\alpha=\{z,x\}$. $\mathbf{S}{}_{i\mu x} = \frac{1}{2} c^{\dag}_{i\mu x\sigma} \left[\mathbf{\sigma}\right]_{\sigma\sigma^\prime} c_{i\mu x\sigma^\prime}$ is the spin operator with Pauli matrix $\mathbf{\sigma}=\left(\sigma_x,\sigma_y,\sigma_z\right)$. $\epsilon_z$ and $\epsilon_x$ represent the on-site energies of the $d_{z^2}$ and $d_{x^2-y^2}$ orbitals respectively. $\hat{P}$ is a projection operator projecting out the double occupancy in the same orbital of all sites.

We adopt the 
TB parameters reported in Ref. \cite{Daoxin_Yao2025}, i.e. $t_{\parallel}=0.445\text{ eV}$, $t_{xz}=0.221\text{ eV}$, $t_{\perp}=0.503\text{ eV}$ and $\epsilon_x-\epsilon_z=0.367\text{ eV}$%
. 
The spin couplings are estimated as $J_{\parallel(\perp)}=\frac{4t_{\parallel(\perp)}^2}{U}$ with $U=10t_{\parallel}$ and $\tilde{J}_\perp=0.5J_\perp$. 
To limit the particle numbers of the two $E_g$ orbitals, we consider the Hamiltonian with Lagrange multipliers
\begin{equation}\label{ap_eq_tildeH2}
    \tilde{H} =\ H + \left(\epsilon^{\prime}_{z}-\epsilon_{z}\right)\sum_{i\mu\sigma}n_{i\mu z\sigma} + \left(\epsilon^{\prime}_{x}-\epsilon_{x}\right)\sum_{i\mu\sigma}n_{i\mu x\sigma},
\end{equation}
where $\epsilon^{\prime}_{z}$ and $\epsilon^{\prime}_{x}$ are determined by $n_{bz}=1$ and $\sum_{\mu\alpha}n_{\mu\alpha}=3$.

The two-orbital model (\ref{ap_eq_tildeH2}) is also solved by SBMF theory. The electron operator is decomposed as $c^{\dag}_{i\mu\alpha\sigma}=f^{\dag}_{i\mu\alpha\sigma}b_{i\mu\alpha}$, where $f$ is spinon operator and $b$ is holon operator. Since we have found that $T_{\mathrm{BEC}} \gg T_{\mathrm{pair}}$ in the considered $n_{bx}$ regime and $T_{\mathrm{pair}}$ is proportional to the zero-temperature spinon pairing gap, we can get the critical temperature of superconductivity only by calculating the ground-state spinon pairing gap. Thus we only consider the spinon Hamiltonian at zero temperature. The superexchange term is also decomposed in $\chi-\Delta$ channel. The spinon Hamiltonian is described as
\begin{equation}\label{SM_eq_MF2}
    \begin{aligned}
        H_{\mathrm{spinon}} = &- t_\parallel\sum_{\langle i,j\rangle,\mu}\delta_{\mu x}\left(f^\dag_{i\mu x\sigma}f_{j\mu x\sigma} + \mathrm{h.c.}\right)
        - t_{xz}\sqrt{\delta_{tx}\delta_{tz}} \sum_{\langle i,j\rangle} \left(f^\dag_{itx\sigma}f_{jtz\sigma} + f^\dag_{itz\sigma}f_{jtx\sigma} + \mathrm{h.c.}\right)\\
        &- \frac{3}{8} J_\parallel\sum_{\langle i,j\rangle\mu}\left(\chi_{ij,\mu x}^\dag\left\langle\chi_{\mu x}\right\rangle + \mathrm{h.c.} + \Delta_{ij,\mu x}^\dag\left\langle\Delta_{\mu x}\right\rangle + \mathrm{h.c.} - \left\langle\chi_{\mu x}^\dag\right\rangle\left\langle\chi_{\mu x}\right\rangle - \left\langle\Delta_{\mu x}^\dag\right\rangle\left\langle\Delta_{\mu x}\right\rangle\right)\\
        &-\frac{3}{8} J_\perp \sum_{i} \left(\chi^{\perp\dag}_{iz}\left\langle\chi^\perp_{z}\right\rangle + \mathrm{h.c.} + \Delta^{\perp\dag}_{iz}\left\langle\Delta^\perp_{z}\right\rangle + \mathrm{h.c.} - \left\langle\chi^{\perp\dag}_{z}\right\rangle\left\langle\chi^\perp_{z}\right\rangle - \left\langle\Delta^{\perp\dag}_{z}\right\rangle\left\langle\Delta^\perp_{z}\right\rangle\right)\\
        &- \frac{3}{8} \tilde{J}_\perp \sum_{i} \left(\Delta^{\perp\dag}_{ix}\langle\Delta^\perp_{x}\rangle + \mathrm{h.c.} - \langle\Delta^{\perp\dag}_{x}\rangle\langle\Delta^\perp_{x}\rangle\right) \\
        &+ \epsilon^\prime_z\sum_{i\mu\sigma}n_{i\mu z\sigma} + \epsilon^\prime_x\sum_{i\mu\sigma}n_{i\mu x\sigma} + \frac{\varepsilon}{2} \sum_{i\alpha\sigma} n_{it\alpha\sigma} - \frac{\varepsilon}{2} \sum_{i\alpha\sigma}n_{ib\alpha\sigma},
    \end{aligned}
\end{equation}
where $\delta_{\mu\alpha}=\left\langle{}b_{i\mu\alpha}b^{\dag}_{j\mu\alpha}\right\rangle$ since holon condense at zero temperature. Under the electric field, we have $\delta_{bz}=0$ and $\delta_{tz}$, $\delta_{tx}$ and $\delta_{bx}$ are solved self-consistently. The mean-field order parameters are represented by
\begin{equation}
    \begin{aligned}
        &\chi_{ij,\mu x} = \sum_{\sigma} f_{i\mu x\sigma}^\dag f_{j\mu x\sigma},\ \chi_{\mu x} = \frac{1}{2N}\sum_{\langle i,j\rangle}\chi_{ij,\mu x},\\
        &\chi^{\perp\dag}_{iz} = \sum_\sigma f^\dag_{izt\sigma}f_{izb\sigma},\ \chi^\perp_z = \frac{1}{N}\sum_i\chi^\perp_{iz},\\
        &\Delta^\dag_{ij,\mu\alpha} = f^\dag_{i\mu\alpha\uparrow}f^\dag_{j\mu\alpha\downarrow}-f^\dag_{i\mu\alpha\downarrow}f^\dag_{j\mu\alpha\uparrow},\ 
        \Delta^{\mathbf{x}(\mathbf{y})}_{\mu x} = \frac{1}{2N}\sum_{\mathbf{R}_i-\mathbf{R}_j=\pm\mathbf{x}(\mathbf{y})}\Delta_{ij,\mu x},\\
        &\Delta^{\perp\dag}_{i\alpha} = f^\dag_{it\alpha\uparrow}f^\dag_{ib\alpha\downarrow}-f^\dag_{ib\alpha\downarrow}f^\dag_{it\alpha\uparrow},\ \Delta^\perp_\alpha = \frac{1}{N}\sum_i\Delta^\perp_{i\alpha}.
    \end{aligned}
\end{equation}

Notably, the spin-exchange $\tilde{J}_\perp$ of the Hamiltonian in Eq. (\ref{ap_eq_H2}) doesn't produce a hopping term $\chi_{x}^\perp$ in Eq. (\ref{SM_eq_MF2}), which is the feature of such a bilayer system. Without interlayer hopping, a small interlayer spin-exchange $J_\perp$ leads to $\langle\chi^\perp\rangle\approx0$.

Consequently, the $3d_{z^2}$ orbital only participate in the interlayer pairing. However, this pairing is not superconductivity (SC) as the corresponding SC order parameter goes to zero in the SBMF theory due to $\delta_{bz}=0$. The SC is carried by the $3d_{x^2-y^2}$ orbitals, which can form both intralayer and interlayer pairing. The superconducting $T_c$ scales with the ground state gap amplitude of the $3d_{x^2-y^2}$ orbitals via the Bardeen-Cooper-Schrieffer (BCS) relation exhibited in Fig.~3(b) in the main text.

The expectation value of the mean-field order parameters are obtained by numerically solving the following self-consistent equations
\begin{equation}
\begin{aligned}
    &\delta_{\mu\alpha} = 1-\frac{1}{N}\sum_k \left(\langle f^\dag_{k\mu\alpha\uparrow}f_{k\mu\alpha\uparrow}\rangle + \langle f^\dag_{-k\mu\alpha\downarrow}f_{-k\mu\alpha\downarrow}\rangle\right),\ \delta_{tz} = 0,\ \sum_{\mu\alpha}\delta_{\mu\alpha} = 1,\\
    &\langle\chi_{\mu x}\rangle = \frac{1}{N}\sum_{\langle i,j\rangle}\langle\chi_{ij,\mu x}\rangle = \frac{1}{N}\sum_k \frac{\cos{(k_x)}+\cos{(k_y)}}{2}\left(\langle f^\dag_{k\mu x\uparrow}f_{k\mu x\uparrow}\rangle + \langle f^\dag_{-k\mu x\downarrow}f_{-k\mu x\downarrow}\rangle\right),\\
    &\langle\chi^\perp_z\rangle = \frac{1}{N}\sum_i\langle\chi^\perp_{iz}\rangle = \frac{1}{N}\sum_k\left(\langle f^\dag_{ktz\uparrow}f_{kbz\uparrow}\rangle + \langle f^\dag_{-ktz\downarrow}f_{-kbz\downarrow}\rangle\right),\\
    &\langle\Delta^{\mathbf{x}(\mathbf{y})}_{\mu x}\rangle^* = \frac{1}{2N}\sum_{\mathbf{R}_i-\mathbf{R}_j=\pm\mathbf{x}(\mathbf{y})}\langle\Delta_{ij,\mu x}\rangle^* = \frac{1}{N}\sum_{k}2\cos{(k_{x(y)})}\langle f^\dag_{k\mu x\uparrow}f^\dag_{-k\mu x\downarrow}\rangle,\\
    &\langle\Delta^\perp_\alpha\rangle^* = \frac{1}{N}\sum_i\langle\Delta^\perp_{i\alpha}\rangle^* = \frac{2}{N}\sum_k \langle f^\dag_{kt\alpha\uparrow}f^\dag_{-kb\alpha\downarrow}\rangle.
\end{aligned}
\end{equation}
Here, $f_{\mathbf{k}\mu\alpha\sigma}=\frac{1}{\sqrt{N}}\sum_{i}f_{i\mu\alpha\sigma}\mathrm{e}^{-\mathrm{i}\mathbf{k}\cdot\mathbf{R}_i}$, $k_{x/y}$ is the $x/y$-component of $\mathbf{k}$, and $\left\langle\Delta_{\mu{}x}^{\mathbf{x}/\mathbf{y}}\right\rangle$ is the $d_{x^2-y^2}$-electron-pairing gap of the nearest neighbor bond along the $x/y$ direction. 

In the self consistent equations, the expectation values of the operators are obtained as follows. The matrix form of the spinon Hamiltonian is 
\begin{equation}
    H_{\mathrm{spinon}} = \sum_{k}
    \left(
    \begin{array}{cccccccc}
        f^{\dagger}_{ktz\uparrow} & f^{\dagger}_{ktx\uparrow} & f^{\dagger}_{kbz\uparrow} & f^{\dagger}_{kbx\uparrow} & f_{-ktz\downarrow} & f_{-ktx\downarrow} & f_{-kbz\downarrow} & f_{-kbx\downarrow}
    \end{array}
    \right)
    \left(
    \begin{array}{cc}
        H_{\chi}(\mathbf{k}) & H_{\Delta}(\mathbf{k}) \\
        H_{\Delta}^{\dagger}(\mathbf{k}) & -H_{\chi}^T(\mathbf{k})
    \end{array}
    \right)
    \left(
    \begin{array}{c}
        f_{ktz\uparrow} \\ f_{ktx\uparrow} \\ f_{kbz\downarrow} \\ f_{kbx\downarrow} \\ f^{\dagger}_{-ktz\downarrow} \\ f_{-ktx\downarrow} \\ f^{\dagger}_{-kbz\downarrow} \\ f_{-kbx\downarrow}
    \end{array}
    \right), 
\end{equation}
where 
\begin{equation}
    \begin{aligned}
        & H_{\chi}(\mathbf{k}) = 
        \left(
        \begin{array}{cccc}
            h_{11}(\mathbf{k}) & h_{12}(\mathbf{k}) & h_{13}(\mathbf{k}) & h_{14}(\mathbf{k}) \\
            h_{21}(\mathbf{k}) & h_{22}(\mathbf{k}) & h_{23}(\mathbf{k}) & h_{24}(\mathbf{k}) \\
            h_{31}(\mathbf{k}) & h_{32}(\mathbf{k}) & h_{33}(\mathbf{k}) & h_{34}(\mathbf{k}) \\
            h_{41}(\mathbf{k}) & h_{42}(\mathbf{k}) & h_{43}(\mathbf{k}) & h_{44}(\mathbf{k})
        \end{array}
        \right), \\
        & h_{11}(\mathbf{k}) = \epsilon_{z}^{\prime} + \frac{\varepsilon}{2}, \ h_{33}(\mathbf{k}) = \epsilon_{z}^{\prime} - \frac{\varepsilon}{2}, \ h_{13}(\mathbf{k}) = h_{31}(\mathbf{k}) = -t_{\perp}\sqrt{\delta_{tz}\delta_{bz}}-\frac{3}8{J_{\perp}}\left\langle\chi^{\perp}_z\right\rangle, \\
        & h_{22}(\mathbf{k}) = -2 \left( t_{\parallel}\delta_{tx} + \frac{3}{8}J_{\parallel}\left\langle\chi_{tx}\right\rangle \right) \left( \cos(k_x) + \cos(k_y) \right) + \epsilon_{x}^{\prime}+\frac{\varepsilon}{2}, \\
        & h_{44}(\mathbf{k}) = -2 \left( t_{\parallel}\delta_{bx} + \frac{3}{8}J_{\parallel}\left\langle\chi_{bx}\right\rangle \right) \left( \cos(k_x) + \cos(k_y) \right) + \epsilon_{x}^{\prime}-\frac{\varepsilon}{2}, \\
        & h_{12}(\mathbf{k}) = h_{21}(\mathbf{k}) = -2t_{xz}\sqrt{\delta_{tz}\delta_{tx}}\left(\cos(k_x)-\cos(k_y)\right), \ h_{34}(\mathbf{k}) = h_{43}(\mathbf{k}) = -2t_{xz}\sqrt{\delta_{bz}\delta_{bx}}\left(\cos(k_x)-\cos(k_y)\right), \\
        & h_{14}(\mathbf{k}) = h_{41}(\mathbf{k}) = h_{23}(\mathbf{k}) = h_{32}(\mathbf{k}) = h_{24}(\mathbf{k}) = h_{42}(\mathbf{k}) = 0, \\
        & H_{\Delta}(\mathbf{k}) = 
        \left(
        \begin{array}{cccc}
            0 & 0 & -\frac{3}{8}J_{\perp}\left\langle\Delta^{\perp}_z\right\rangle & 0 \\
            0 & -\frac{3}{4}J_{\parallel} \left( \left\langle\Delta_{tx}^{\mathbf{x}}\right\rangle\cos(k_x) + \left\langle\Delta_{tx}^{\mathbf{y}}\right\rangle\cos(k_y) \right) & 0 & -\frac{3}{8}\tilde{J}_{\perp} \left\langle\Delta^{\perp}_x\right\rangle \\
            -\frac{3}{8}J_{\perp}\left\langle\Delta^{\perp}_z\right\rangle & 0 & 0 & 0 \\
            0 & -\frac{3}{8}\tilde{J}_{\perp} \left\langle\Delta^{\perp}\right\rangle & 0 & -\frac{3}{4}J_{\parallel} \left( \left\langle\Delta_{bx}^{\mathbf{x}}\right\rangle\cos(k_x) + \left\langle\Delta_{bx}^{\mathbf{y}}\right\rangle\cos(k_y) \right)
        \end{array}
        \right). 
    \end{aligned}
\end{equation}

$H_{\mathrm{spinon}}$ can be diagonalized by a Bogoliubov transformation. 
\begin{equation}
    H_{\mathrm{spinon}} = \sum_{k} 
    \left(
    \begin{array}{ccc}
        \gamma^{\dagger}_{k1} & \cdots & \gamma^{\dagger}_{k8}
    \end{array}
    \right) \ 
    \mathrm{diag}(E_{k1},\cdots,E_{k8}) \ 
    \left(
    \begin{array}{c}
        \gamma_{k1} \\ \vdots \\ \gamma_{k8}
    \end{array}
    \right), 
\end{equation}
where
\begin{equation}
    \xi^{\dagger}(\mathbf{k}) 
    \left(
    \begin{array}{cc}
        H_{\chi}(\mathbf{k}) & H_{\Delta}(\mathbf{k}) \\
        H_{\Delta}^{\dagger}(\mathbf{k}) & -H_{\chi}^T(\mathbf{k})
    \end{array}
    \right)
    \xi(\mathbf{k}) 
    = \mathrm{diag}(E_{k1},\cdots,E_{k8}), \ 
    \left(
    \begin{array}{c}
        f_{ktz\uparrow} \\ f_{ktx\uparrow} \\ f_{kbz\uparrow} \\ f_{kbx\uparrow} \\ f^{\dagger}_{-ktz\downarrow} \\ f^{\dagger}_{-ktx\downarrow} \\ f^{\dagger}_{-kbz\downarrow} \\ f^{\dagger}_{-kbx\downarrow}
    \end{array}
    \right) 
    = \xi(\mathbf{k}) 
    \left(
    \begin{array}{c}
        \gamma_{k1} \\ \vdots \\ \gamma_{k8}
    \end{array}
    \right). 
\end{equation}
Here, $\gamma$ is fermionic operator, whose expectation value is 
\begin{equation}
    \left\langle\gamma^{\dagger}_{km}\gamma_{k^{\prime}n}\right\rangle = \delta_{kk^{\prime}}\delta_{mn}n_F(E_{km}), \ 
    \left\langle\gamma_{km}\gamma_{k^{\prime}n}\right\rangle = \left\langle\gamma^{\dagger}_{km}\gamma^{\dagger}_{k^{\prime}n}\right\rangle = 0, 
\end{equation}
where $m=1,\cdots,8$ and $n_F(E_{km})=1/\left(\mathrm{e}^{E_{km}/k_BT}+1\right)$ is Fermi distribution function with $T$ representing the temperature. Then the order parameter operators are expressed as 
\begin{equation} \label{ap_SC2_T}
    \begin{aligned}
        & \delta_{tz(tx,bz,bx)} = 1 - \frac{1}{N} \sum_{k} \sum_{m=1,\cdots,8} \left[ \xi^{\ast}_{1(2,3,4),m}(\mathbf{k})\xi_{1(2,3,4),m}(\mathbf{k})n_F(E_{km}) + \xi^{\ast}_{5(6,7,8),m}(\mathbf{k})\xi_{5(6,7,8),m}(\mathbf{k})\left(1-n_F(E_{km})\right) \right], \\
        & \left\langle\chi_{tx(bx)}\right\rangle = \frac{1}{N} \sum_{k} \frac{\cos(k_x)+\cos(k_y)}{2} \sum_{m=1,\cdots,8} \left[ \xi^{\ast}_{2(4),m}(\mathbf{k})\xi_{2(4),m}(\mathbf{k})n_F(E_{km}) + \xi^{\ast}_{6(8),m}(\mathbf{k})\xi_{6(8),m}(\mathbf{k})\left(1-n_F(E_{km})\right) \right], \\
        & \left\langle\chi^{\perp}_z\right\rangle = \frac{1}{N} \sum_{k} \sum_{m=1,\cdots,8} \left[ \xi^{\ast}_{1,m}(\mathbf{k})\xi_{3,m}(\mathbf{k})n_F(E_{km}) + \xi^{\ast}_{7,m}(\mathbf{k})\xi_{5,m}(\mathbf{k})\left(1-n_F(E_{km})\right) \right], \\
        & \left\langle\Delta_{tx(bx)}^{\mathbf{x}/\mathbf{y}}\right\rangle^{\ast} = \frac{1}{N} \sum_{k} 2\cos(k_{x/y}) \sum_{m=1,\cdots,8} \xi^{\ast}_{2(4),m}(\mathbf{k})\xi_{6(8),m}(\mathbf{k})n_F(E_{km}), \\
        & \left\langle\Delta^{\perp}_{z(x)}\right\rangle = \frac{2}{N} \sum_{k} \sum_{m=1,\cdots,8} \xi^{\ast}_{1(2),m}(\mathbf{k})\xi_{7(8),m}(\mathbf{k})n_F(E_{km}),
    \end{aligned}
\end{equation}
where $\xi_{a,m}$ represents the matrix element of $\xi$ at the $a$-th row and the $m$-th column. 
The method to solve these self-consistent equations is similar to the one to solve the one-orbital model. 

In the two-orbital model, the pairing gap amplitude $\tilde{\Delta}_x$ is defined as the maximal value of the pairing gap of the $3d_{x^2-y^2}$ electrons. For the interlayer $s$-wave pairing, $\tilde{\Delta}_{x}=-\frac{3}{8}\tilde{J}_{\perp}\left\langle\Delta^{\perp}_{x}\right\rangle$. For the ($s$($d_{z^2}$)$+$i$d$($d_{x^2-y^2}$))-wave pairing, $\tilde{\Delta}_{x}=-\frac{3}{2}J_{\parallel}\left\langle\Delta^{\mathbf{x}}_{bx}\right\rangle$.

\begin{figure}[htbp]
    \centering
    \includegraphics[width=0.36\linewidth]{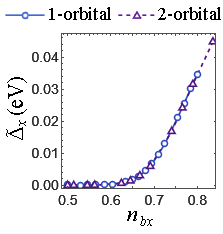}
    \caption{The comparison of the pairing amplitude $\tilde{\Delta}_{x}$ as function of the bottom-layer particle number $n_{bx}$ calculated with different models. The lines almost coincide with each other, particularly in the large-$n_{bx}$ regime. } \label{fig_Deltad_1orb2orb}
\end{figure}

We find that the pairing amplitude $\tilde{\Delta}_{x}$ as function of the bottom-layer particle number $n_{bx}$ calculated with different models show a strong resemblance (see Fig. \ref{fig_Deltad_1orb2orb}).

\section{C. More numerical simulation results by density matrix renormalization group} 
\label{appendix_c}
We apply the density matrix renormalization group (DMRG) approach \cite{white1993dmrg,weng1999dmrg} to solve the ground state of Hamiltonian (\ref{ap_eq_H1}) with $t_{\parallel}=1$, $J_{\parallel}=0.4t_{\parallel}$ and $\tilde{J}_\perp=0.8J_{\parallel}$. The definitions of all the operators and main results have been shown in the main text. Here we give more results.

\begin{figure}[htbp]
    \centering
    \includegraphics[width=0.72\linewidth]{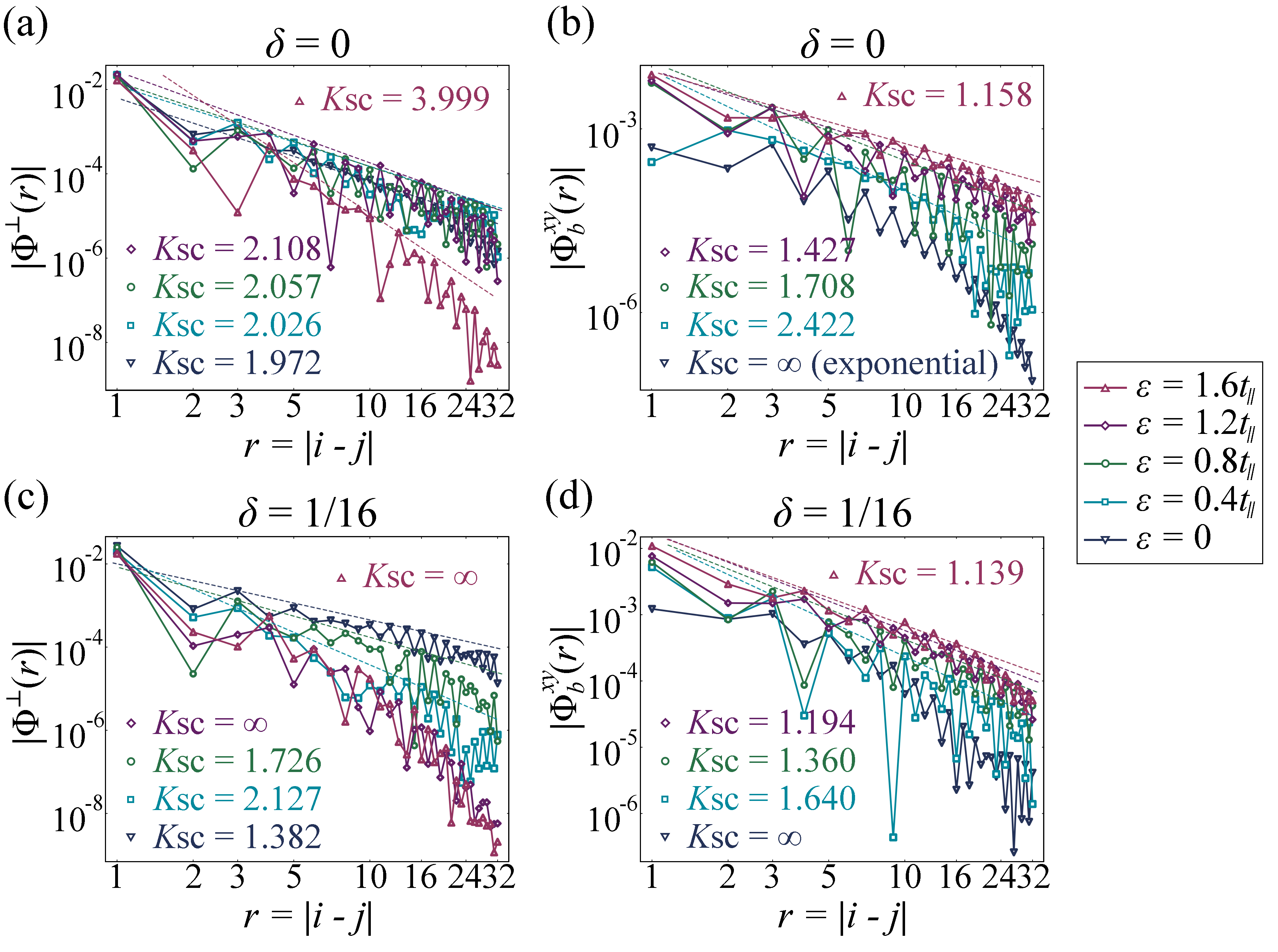}
    \caption{(a)-(d) The interlayer pairing correlation functions $\left|\Phi^{\perp}(r)\right|$ and intralayer correlation functions $\left|\Phi^{\mathbf{xy}}_b(r)\right|$ under different electronic field $\varepsilon = 0, 0.4t_\parallel,0.8t_\parallel,1.2t_\parallel,1.6t_\parallel$ for $\delta=0$ in (a)-(b) and $\delta=1/16$ in (c)-(d). The decay exponents $K_\text{SC}$ are written in the figures as well.} \label{fig_dmrg_phi}
\end{figure}

In Fig. \ref{fig_dmrg_phi}, we show the absolute value of correlation functions $\left|\Phi^{\perp}(r)\right|$ and $\left|\Phi^{\mathbf{xy}}_t(r)\right|$ for comparison with Fig. 4 in the main text. It turns out that the external electric field suppresses interlayer pairing across different electron-doped levels, while it notably strengthens intra-bottom-layer pairing by increasing carrier concentration.

\begin{figure}[htbp]
    \centering
    \includegraphics[width=0.6\linewidth]{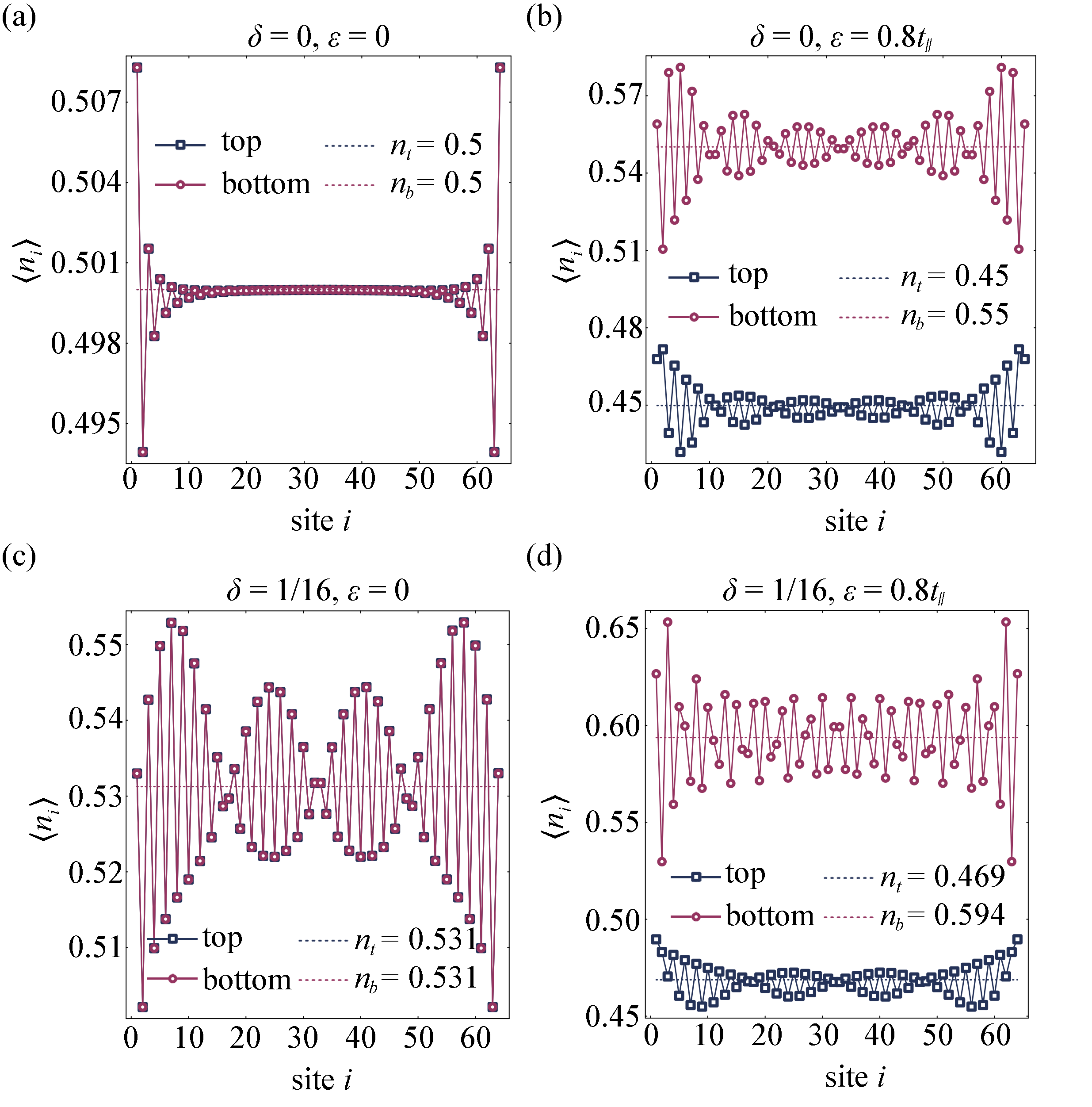}
    \caption{(a)-(d) The particle number distribution in the two layers for different parameters $(\delta,\varepsilon)=(0,0)$, $(0,0.8t_\parallel)$, $(1/16,0)$ and $(1/16,0.8t_\parallel)$. The dashed lines correspond to the total expectation of the number of electrons in each of the two layers.} \label{fig_dmrg_n}
\end{figure}

Fig. \ref{fig_dmrg_n} shows the particle number distribution in the two layers for different $(\delta,\varepsilon)$. As shown in Fig. \ref{fig_dmrg_n} (a) and (c), the expectation of electron number in the two layers is equal without an external electric field. Fig. \ref{fig_dmrg_n} (b) and (d) exhibit that when a perpendicular electric field is introduced, electrons flow from the top layer with a lower electric potential to the bottom layer with a higher electric potential.

\begin{figure}[htbp]
    \centering
    \includegraphics[width=0.6\linewidth]{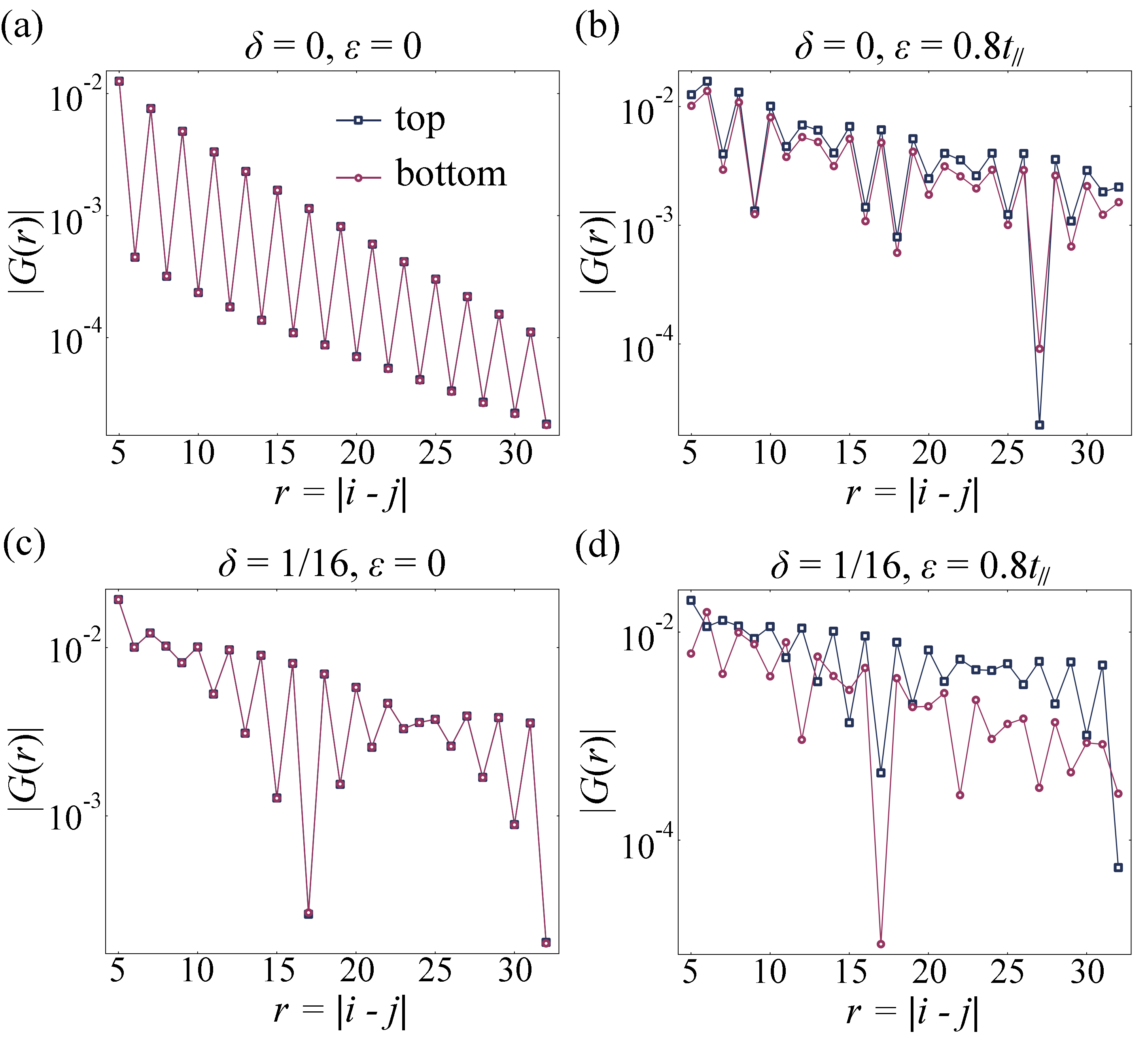}
    \caption{(a)-(d) The single particle Green's function $|G(r)|$ for $(\delta,\varepsilon)=(0,0)$, $(0,0.8t_\parallel)$, $(1/16,0)$ and $(1/16,0.8t_\parallel)$.} \label{fig_dmrg_g}
\end{figure}

We also consider the single particle Green's function, defined as
\begin{equation}
    G_{\mu}(r) = \sum_{\sigma} \left\langle{}c^{\dag}_{i\mu\sigma}c_{j\mu\sigma}\right\rangle\ (\mu=t,b).
\end{equation}
The results are shown in Fig. \ref{fig_dmrg_g}. The Green's function under all of the parameters exhibits exponential decay. 

\section{D. The effect of interlayer coulomb interaction} 
\label{appendix_d}
In order to better capture the properties of La$_3$Ni$_2$O$_7$ thin films, we further consider the effect of interlayer Coulomb interaction in the single-orbital model and solve it using both SBMF and DMRG methods, respectively. The modified Hamiltonian is
\begin{equation}\label{ap_eq_addV}
    H=-t_{\parallel} \sum_{\left\langle{}i,j\right\rangle,\mu,\sigma} \hat{P} \left( c^{\dag}_{i\mu\sigma}c_{j\mu\sigma} + \mathrm{h.c.} \right) \hat{P} + \sum_{i,\mu} \epsilon_{\mu} n_{i\mu} + J_{\parallel} \sum_{\left\langle{}i,j\right\rangle,\mu} \mathbf{S}_{i\mu} \cdot \mathbf{S}_{j\mu} +  \tilde{J}_{\perp} \sum_{i} \mathbf{S}_{it} \cdot \mathbf{S}_{ib} + V\sum_i n_{it}n_{ib},
 \end{equation}
where $V$ is the strength of the interlayer Coulomb interaction.

In the SBMF method, the additional terms are decomposed as
\begin{equation}\label{ap_eq_decom}
\begin{aligned}
    V\sum_i n_{it}n_{ib} \rightarrow
    & V\sum_i (\langle n_{t}\rangle n_{ib} + \langle n_{b}\rangle n_{it} - \langle n_{t}\rangle\langle n_{b}\rangle)\\
    & - \frac{1}{2} V \sum_{i} \left(\chi^{\perp\dag}_{iz}\left\langle\chi^\perp_{z}\right\rangle + \mathrm{h.c.} - \left\langle\chi^{\perp\dag}_{z}\right\rangle\left\langle\chi^\perp_{z}\right\rangle\right)\\
    & + \frac{1}{2} V \sum_{i} \left(\Delta^{\perp\dag}_{iz}\left\langle\Delta^\perp_{z}\right\rangle + \mathrm{h.c.} - \left\langle\Delta^{\perp\dag}_{z}\right\rangle\left\langle\Delta^\perp_{z}\right\rangle\right),
\end{aligned}
\end{equation}
where $\langle n_\mu\rangle = n_\mu$ is average particle number expectation for layer $\mu=t, b$. 
The expectation value of the first term is
\begin{equation}
    V N \left\langle n_{t} \right\rangle \left\langle n_b \right\rangle = \frac{VN}{4} \left( \left\langle n_t \right\rangle + \left\langle n_b \right\rangle \right)^2 - \frac{VN}{4} \left( \left\langle n_b \right\rangle - \left\langle n_t \right\rangle \right)^2, 
\end{equation}
indicating that the $V$ term favors the difference of the particle number of different layers by lower the energy. 
The expectation value of the last term is 
\begin{equation}
    +\frac{VN}{2} \left| \left\langle \Delta^{\perp} \right\rangle \right|^2, 
\end{equation}
therefore the $V$ term is expected to suppress the interlayer pairing by lifting its energy. 

\begin{figure}[htbp]
    \centering
    \includegraphics[width=0.4\linewidth]{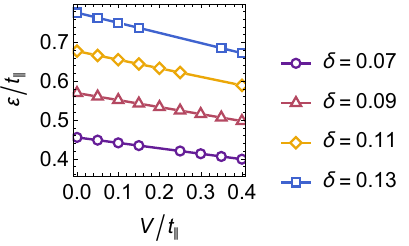}
    \caption{The relationship between external electric field $\varepsilon$ and interlayer Coulomb interaction $V$ for realizing the same particle number distribution ($n_t=0.5-\delta/2$ and $n_b=0.5+3\delta/2$) and the same $d$-wave pairing gap (obtained by SBMF approach). } \label{fig_interCoulomb_1orb_SBMF}
\end{figure}

\begin{figure}[htbp]
    \centering
    \includegraphics[width=0.75\linewidth]{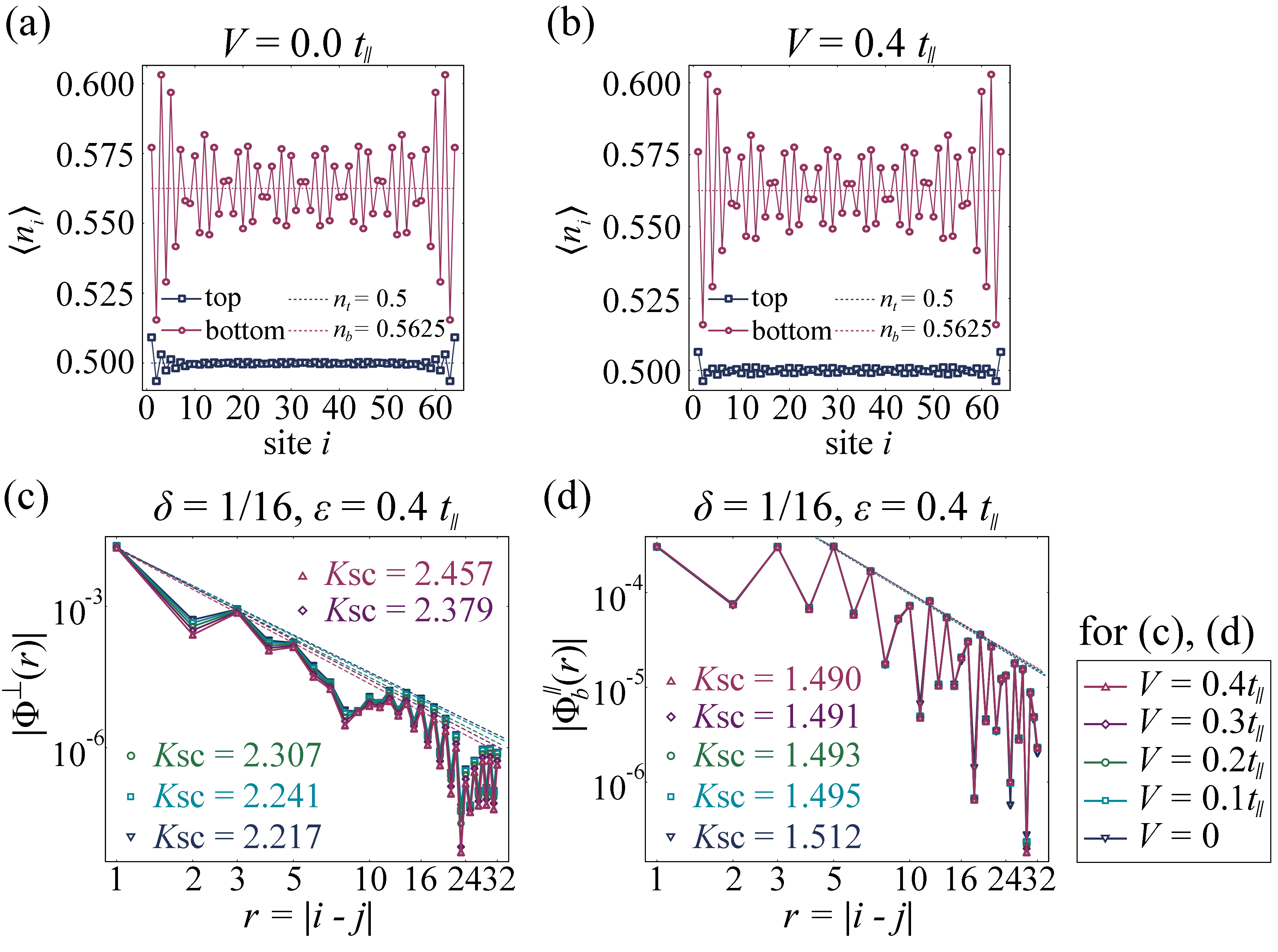}
    \caption{DMRG results for the Hamiltonian Eq. (\ref{ap_eq_addV}) including the interlayer Coulomb interaction $V$-term. (a) The interlayer pairing correlation functions $|\Phi^\perp(r)|$ for different $V$ within $V=0\sim0.4t_\parallel$. (b) The intra-bottom-layer pairing correlation functions $|\Phi^\parallel_b(r)|$ for different $V$ within $V=0\sim0.4t_\parallel$. The algebraic decay coefficient $K_\text{SC}$ reflects the decay rate of the pairing correlation function with spatial distance, negatively correlated with the corresponding pairing strength. } \label{fig_DMRG_V}
\end{figure}

Fig. \ref{fig_interCoulomb_1orb_SBMF} shows the SBMF results, exhibiting the relationship between external electric field $\varepsilon$ and the interlayer Coulomb interaction $V$ with the same particle number distribution and the same $d$-wave pairing gap. The results display that, when realizing the same particle number distribution ($n_t=0.5-\delta/2$ and $n_b=0.5+3\delta/2$) and the same pairing gap, the external electric field $\varepsilon$ sightly decreases as the interlayer Coulomb interaction $V$ increases, indicating that interlayer Coulomb interaction slightly promotes charge transfer between layers. Therefore, if the external electric field $\varepsilon$ is fixed, introducing interlayer Coulomb interaction will slightly increase the difference of the particle number of the two layers as well as the particle number of the bottom layer, and thus will increase the critical temperature of $d$-wave pairing, which only depends on the particle number of the bottom layer.

The DMRG results shown in Fig. \ref{fig_DMRG_V} illustrates the trend of the properties of ground state with different interlayer Coulomb interactions $V = 0 \sim 0.4t_\parallel$. Here we take the transferred $d_{x^2-y^2}$-electron-doping level $\delta = 1/16$ and $\varepsilon = 0.4t_\parallel$ for a typical analysis, and the other parameters are consistent with the main text. Fig. \ref{fig_DMRG_V} (a) and (b) show the particle number distributions for $V = 0$ and $0.4t_\parallel$, and the results indicate that the interlayer Coulomb interaction does not have an observable effect on the particle number distributions, which may also be affected by the finite size. Fig. \ref{fig_DMRG_V} (c) and (d) exhibit the trends of interlayer superconducting pairing and intralayer pairing with different $V$, respectively. According to the trend of the algebraic decay coefficient $K_\text{SC}$ shown in Fig. \ref{fig_DMRG_V}, the interlayer pairing is suppressed with the enhancement of $V$, while the intra-bottom-layer pairing demonstrates slight enhancement, which is qualitatively consistent with the SBMF results.

In conclusion, the combined SBMF and DMRG results show that the introduction of interlayer Coulomb interaction slightly increases the particle number difference between the two layers and the strength of intra-bottom-layer pairing, while suppressing interlayer pairing.



\newpage

\twocolumngrid

\bibliography{reference}

\end{document}